\documentclass[journal=jpca,manuscript=article,monochrome]{achemso}
\setkeys{acs}{articletitle=true}

\usepackage{chemformula} 
\usepackage[T1]{fontenc} 
\SectionNumbersOn

\usepackage{bm}
\usepackage{graphicx}
\usepackage{dcolumn}
\usepackage{bm}
\usepackage{array}
\usepackage{multirow}
\usepackage{url}
\usepackage{tabularx}
\usepackage{braket}
\usepackage[bookmarks=true,colorlinks=true,urlcolor=blue,linkcolor=blue,citecolor=blue]{hyperref}
\usepackage{longtable}
\usepackage{makecell}
\author{Rebekah Hermsmeier$^{1}$, Xiaodong Xing$^{1}$, and Timur V. Tscherbul$^{1}$}
\affiliation{$^{1}$Department of Physics, University of Nevada, Reno, NV 89557, USA}
\email{ttscherbul@unr.edu}

\title{Nuclear spin relaxation in cold atom-molecule collisions}

\date{\today}

\begin{document}

\begin{tocentry}
\includegraphics[scale=0.22,  trim = 0 10 0 0]{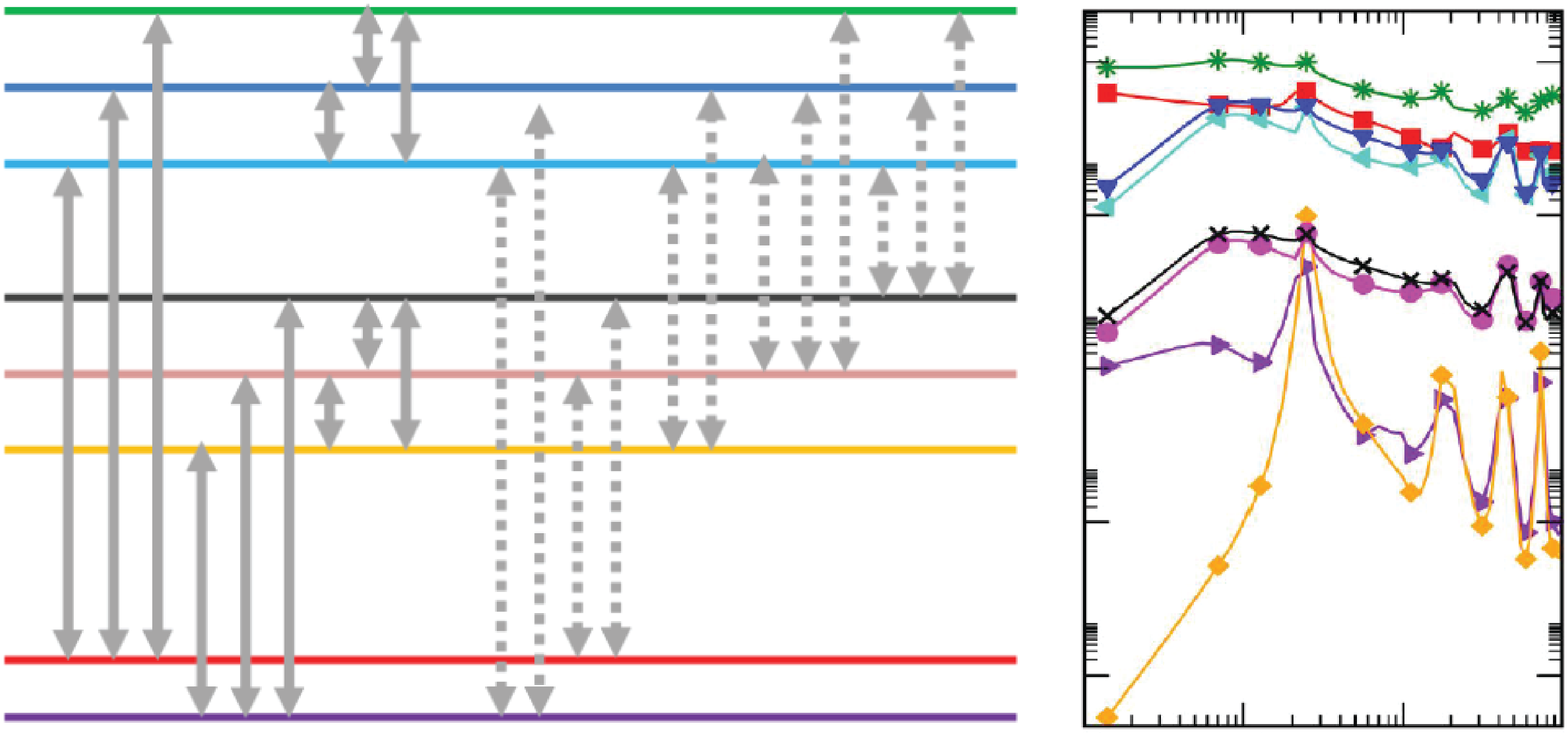} 
\end{tocentry}

\begin{abstract} 

We explore the quantum dynamics of nuclear spin relaxation in cold collisions of $^1\Sigma^+$ molecules with structureless atoms in an external  magnetic field. To this end, we develop a rigorous coupled-channel methodology, which accounts for rotational and nuclear spin degrees of freedom of  $^1\Sigma^+$ molecules, their interaction with an external magnetic field, as well as for anisotropic atom-molecule interactions. We apply the methodology to study collisional relaxation of the nuclear spin sublevels  of $^{13}$CO  molecules immersed in a cold buffer gas of $^4$He atoms. We find that nuclear spin relaxation in the ground rotational manifold ($N=0$) of $^{13}$CO occurs extremely slowly due to the absence of direct couplings between the nuclear spin sublevels. The rates of collisional transitions between the rotationally excited ($N=1$) nuclear spin states of $^{13}$CO  are generally much higher due to the direct nuclear spin-rotation coupling between the states. These transitions obey selection rules, which depend on the values of space-fixed projections of rotational and nuclear spin angular momenta ($M_N$ and $M_I$) for the initial and final molecular states.   For some initial states, we also observe a strong  magnetic field dependence, which can be understood using the first Born approximation.  We use our calculated nuclear spin relaxation rates to investigate the thermalization of a single nuclear spin state of $^{13}$CO$(N=0)$ immersed in a cold  buffer gas of $^4$He. The calculated nuclear spin relaxation times ($T_1\simeq 1$~s at $T=1$~K at the He density of 10$^{-14}$ cm$^{-3}$) display a steep temperature dependence decreasing rapidly at elevated temperatures due to  the increased population of rotationally excited states, which undergo nuclear spin relaxation at a much faster rate.
Thus, long relaxation times of $N=0$ nuclear spin states in cold collisions with buffer gas atoms can only be maintained  at sufficiently low temperatures \textcolor{red}{($k_BT\ll 2B_e$)}, where $B_e$ is the rotational constant.

\color{black}



\end{abstract}

\maketitle
\section{Introduction}

Cold and ultracold molecular gases prepared in single rovibrational and spin quantum states can be efficiently controlled with external electromagnetic fields \cite{Carr:09,Lemeshko:13}, thereby forming a unique platform for exploring fundamental concepts of  gas-phase reaction dynamics, such as long-lived complex formation, universal dynamics, external field control, and the role of quantum chaos in chemical reactivity \cite{Krems:08,Balakrishnan:16,Bohn:17}.
Ultracold polar molecules also hold promise for quantum information science, precision spectroscopy, and searches for new physics beyond the Standard Model \cite{Carr:09,Bohn:17,Hutzler:20}.  Experimental realization of these proposals demands dense, cold, and long-lived molecular ensembles. As such, understanding low-temperature collisions within these ensembles, which limit both the maximum achievable density and lifetime,  has long been a major thrust in the field \cite{Krems:08,Balakrishnan:16,Bohn:17}.

Atom-molecule and molecule-molecule collisions can have beneficial as well as detrimental effects on the stability of cold molecular gases. While elastic collisions are a main driving force behind   sympathetic  and evaporative cooling \cite{Lara:06,Carr:09,Tscherbul:11}  
inelastic collisions lead to heating and/or trap loss. In particular, inelastic collisions can flip the orientation of molecular electron spins,   leading to spin relaxation (also known as spin depolarization) \cite{Volpi:02,Krems:04,Tscherbul:11}. As collisional spin relaxation is a primary loss mechanism for  magnetically trapped molecules, it has been the subject of much experimental and theoretical work (see, e.g., Refs.~\cite{Weinstein:98,Campbell:09,Hummon:11,Singh:13,Segev:19,Son:20,Jurgilas:21,Volpi:02,Krems:04,Tscherbul:11,Suleimanov:12,Morita:18,Morita:19b} and references therein).
 Volpi and Bohn \cite{Volpi:02} and Krems and Dalgarno \cite{Krems:04} performed the first rigorous coupled-channel calculations of cold collisions between open-shell molecules and atoms in the presence of an external magnetic field. 
This work has since been extended to a variety of ultracold atom-molecule and molecule-molecule systems, and generated theoretical predictions of their low-temperature collisional properties \cite{Tscherbul:11,Morita:18,Morita:19b,Suleimanov:12,Tscherbul:15}. 


By comparison, collisional relaxation of {\it nuclear} spins has drawn much less attention.
   Nuclear spin-flipping collisions are responsible for the stability of nuclear spin states of molecules immersed in cold inert buffer gases (such as He or Ne). These systems can be realized experimentally using cryogenic buffer-gas cooling \cite{Weinstein:98, Patterson:09,Hutzler:12,Porterfield:19,Egorov:01,Lu:09,Patterson:13,Iwata:17,Changala:19,Santamaria:21,Hofsass:21,Daniel:21,Liu:22,Bu:22}, and they are  interesting for a variety of  reasons. First, preparing molecules in a single nuclear spin (or hyperfine) state  enhances the sensitivity of   spectroscopic measurements \cite{Egorov:01,Lu:09,Patterson:13,Iwata:17,Changala:19,Santamaria:21,Hofsass:21,Daniel:21,Liu:22,Bu:22} and is essential for the initialization steps of molecule-based quantum information processing protocols. 
    One example is  hyperpolarized nuclear magnetic resonance (NMR), which relies on driving populations of nuclear spin states out of thermal equilibrium as a means to enhance the sensitivity of conventional (thermal) NMR \cite{Walker:97,Gentile:17,Barskiy:17,Spiliotis:21}.
Because nuclear spins  interact weakly with their environment, they could be an ideal platform for long-term quantum information storage \cite{Park:17,Gregory:21}.  Our ability to use buffer gas-cooled molecules for these applications is currently hindered by the lack of knowledge of collisional nuclear spin relaxation rates. Indeed, if  these rates turn out to be large, collisional thermalization would lead to rapid decoherence of the nuclear spin superposition states, making them unsuitable for quantum information processing.

Nuclear spins can affect molecular collisions and chemical reactions through several mechanisms.
First, nuclear spin statistics restricts the number of available reactants and/or product states. As a prime example, only odd partial waves are allowed for collisions of identical fermions in the same internal states, leading to a suppression of the ultracold  chemical reaction  KRb~+~KRb $\to$ K$_2$~+~Rb$_2$ \cite{Ospelkaus:10b,Ni:10}. Homonuclear diatomic molecules can exist in the form of different nuclear spin isomers, such as ortho and para-H$_2$, which can  exhibit  dramatically 
different  chemical reactivity  at ultralow temperatures, as seen in theoretical calculations \cite{Balakrishnan:16,Simbotin:15}.  Nuclear spin isomers of polyatomic molecules  such as methylene (CH$_2$) have been predicted to have markedly different spin relaxation rates in cold collisions with He atoms \cite{Tscherbul:12b} and ortho- and para-water molecules have different reactivity towards trapped diazenylium ions \cite{Kilaj:18}.
Second, because nuclear spins are weakly coupled to the other degrees of freedom, it is expected that the total nuclear spin of the collision complex should  be  conserved, which leads to nuclear spin selection rules \cite{oka2004nuclear}. 
These selection rules have recently been observed experimentally for the ultracold chemical reaction KRb~+~KRb $\to$ K$_2$~+~Rb$_2$, which populates only  even (odd) rotational states of K$_2$ (Rb$_2$) \cite{Hu:20} when the reactants are prepared in single, well-defined nuclear spin states.
Finally, hyperfine interactions between the nuclear spins and the other degrees of freedom (such as the electron spins in open-shell atoms and molecules) play a crucial role in low-temperature atomic and molecular collisions  \cite{Lara:07,Tscherbul:07,GonzalezMartinez:11,Tscherbul:20,Hermsmeier:21} being largely responsible for, e.g., the occurrence of magnetic Feshbach resonances in ultracold atom-atom collisions \cite{Chin:10}.
However, rigorous theoretical studies of nuclear spin effects in ultracold molecular collisions have been largely limited to hyperfine interactions in open-shell molecule-atom collisions  \cite{Lara:06,Tscherbul:07,GonzalezMartinez:11}.
Quem\'en\'er {\it et al.} recently proposed a simple state decomposition model \cite{Quemener:21} to describe the  effects of nuclear spin conservation and  external magnetic fields on the product state distributions  of the ultracold chemical reaction KRb~+~KRb $\to$ K$_2$~+~Rb$_2$. While the model describes these effects remarkably well, it makes a number of assumptions, such as neglecting the rotational structure of the reactants and products. Model calculations on ultracold RbCs~+~RbCs  \cite{Wallis:14}, Li~+~CaH \cite{Tscherbul:20}, and Na~+~NaLi collisions \cite{Hermsmeier:21}  used severely limited basis sets, which  did not produce converged results when hyperfine degrees of freedom were included.

Here, we develop a rigorous quantum mechanical theory of nuclear spin relaxation  in collisions of $^1\Sigma$ molecules with structureless atoms in the presence of an external magnetic field. We apply the theory to calculate numerically converged cross sections and rate coefficients for  transitions between the different rotational and nuclear spin sublevels of $^{13}$CO molecules in low-temperature collisions  with $^4$He atoms, and to explore their dependence  on collision energy and magnetic field.  Our calculations show that such transitions follow distinct selection rules.  For example, nuclear spin-flipping transitions occur very slowly in the ground rotational state manifold, leading to  nuclear spin relaxation ($T_1$) times on the order of 0.5 s at the buffer-gas density of 10$^{-14}$ cm$^3$/s and $T=1$~K. The long relaxation times  of  the nuclear spin sublevels of the ground rotational state implies their potential utility for precision spectroscopy and quantum information storage. The long $T_1$ times are maintained as long as the buffer gas temperature is much lower than the spacing between the ground and the first excited rotational levels.

The rest of this paper is structured as follows. In Sec. II we present the quantum scattering methodology for  atom-molecule collisions in a magnetic field, which explicitly includes  the nuclear spin degrees of freedom of $^1\Sigma$ molecules.  We then apply the methodology  to obtain converged cross sections for nuclear spin transitions in cold He~+~CO collisions. The   relevant computational details are given at the end of Sec. II.  In Sec. III, we present and analyze the cross sections and rate constants for nuclear spin relaxation in cold He~+~CO collisions. In Sec. IIID we consider the dynamics of nuclear spin sublevels of CO molecules immersed in a cold buffer gas of He. Section IV summarizes the main results of this work.

\section{Theory}

In this section, we will develop the quantum theory of collisions between $^1\Sigma$ molecules bearing a single nuclear spin (such as $^{13}$C$^{16}$O) and structureless S-state atoms in an external magnetic field. We will next apply the theory to calculate the cross sections and rates for nuclear spin-changing transitions in cold $^4$He~+~$^{13}$C$^{16}$O($^1\Sigma^+$) collisions.  

The Hamiltonian of the atom-molecule collision complex may be  written as
\begin{equation}\label{Hfull}
\begin{aligned}
\hat{H}=-\dfrac{1}{2\mu R}\dfrac{\partial^2}{\partial R^2}R+\dfrac{\hat{L}^2}{2\mu R^2}+\hat{V}(R,r,\theta)+\hat{H}_{\rm mol},
\end{aligned}
\end{equation}
where the orbital angular momentum operator $\hat{L}$ describes the orbital motion of the colliding particles,  $\mu={M_{\rm at}M_{\rm mol}}/(M_{\rm at}+M_{\rm mol})$ is the reduced mass of the complex, and \textcolor{red}{$\hat{V}$ represents the atom-molecule interaction potential in Jacobi coordinates $(R,r,\theta)$}, where $r=|\mathbf{r}|$ is the internuclear distance in the diatomic molecule, $R=|\mathbf{R}|$ is the separation vector from the atom to the center of mass of the molecule,  
and $\theta$ is the angle between $\mathbf{R}$ and $\mathbf{r}$. \textcolor{red}{Here, we consider collisions of $^{13}$C$^{16}$O molecules with structureless atoms (such as $^4$He), and hence the atomic Hamiltonian can be omitted from Eq.~\eqref{Hfull}. The interaction potential $V(R,r,\theta)$ approaches zero as $R \to \infty$.}

The effective Hamiltonian of the $^1\Sigma^+$ molecule in its ground electronic and vibrational states
\cite{Brown:03,meerts1977electric},

\begin{equation}
\begin{aligned}
\hat{H}_{\rm mol}=\hat{H}_{\rm rot}+\hat{H}_{\rm hf}+\hat{H}_{\rm Z}
\end{aligned}
\end{equation}
can be decomposed into the rotational, hyperfine, and Zeeman terms
\begin{equation}
\begin{aligned}\label{Hmol}
\hat{H}_{\rm rot}&=B_e \hat{\mathbf{N}}^2-D_v\hat{\mathbf{N}}^4,\\ 
\hat{H}_{\rm hf}&=A \hat{\mathbf{I}}\cdot \hat{\mathbf{N}},\\
\hat{H}_{\rm Z}=-g_N\mu_N \hat{N}_z B &- g_I\mu_N \hat{I}_z  B - \dfrac{1}{\sqrt{6}}B^2\sum_qD_{0q}^{{2}*}(\omega)\hat{T}_q^2({\chi}),\\
\end{aligned}
\end{equation}
where $B_e$ is the rotational constant, $D_v$ is the centrifugal distortion constant, $\hat{\mathbf{N}}$ is the rotational angular momentum operator, $\hat{\mathbf{I}}$ is the nuclear spin operator, $A$ is the nuclear spin-rotation interaction constant, $g_N$ is the rotational g-factor, $g_I$ is the nuclear g-factor, $\mu_N$ is the nuclear magneton, $B$ is the magnetic field, $D_{0q}^2(\omega)$ is a Wigner $D$-function of the Euler angles $\omega$, which determine the position of the molecular axis in the space-fixed frame, and $\hat{T}_q^2(\chi)$ is the magnetic susceptibility tensor \cite{meerts1977electric}.

The hyperfine structure of $^{13}$CO arises from the nuclear spin of $^{13}$C $(I=1/2)$ and includes the nuclear spin-rotation interaction defined by $\hat{H}_{\rm hf}$ in Eq. (\ref{Hmol}). The Zeeman  term $\hat{H}_{\rm Z}$ accounts for the interaction of the external magnetic field with molecular rotational angular momentum, nuclear spin, and diamagnetic susceptibility  \cite{Brown:03} represented by the first, second, and third terms in the third line of Eq.~(\ref{Hmol}).

We assume that the external magnetic field $B$ is directed along the space-fixed (SF) quantization axis $z$. 
The Hamiltonian in Eq.~(\ref{Hmol}) employs the rigid rotor approximation with a correction for centrifugal distortion. This effectively neglects the vibrational motion of the molecule, which is known to be a good approximation for collisions with weakly perturbing buffer gas atoms (such as $^4$He) at low temperatures \cite{Balakrishnan:16,Balakrishnan2000jcp}.

To solve the quantum  scattering problem for the atom-molecule collision system, we expand the total wavefunction of the system in a complete set of uncoupled basis functions in the SF frame
\begin{equation}
 |\Psi\rangle=\dfrac{1}{R}\sum_{N,M_N}\sum_{M_I}\sum_{L,M_L} F^M_{NM_N IM_I LM_L}(R) |NM_N\rangle |IM_I\rangle |LM_L\rangle,
\label{eq:psi}
\end{equation}
where $M_N$, $M_I$ and $M_L$ indicate the projections of $\hat{\mathbf{N}}$, $\hat{\mathbf{I}}$ and $\hat{\mathbf{L}}$  onto the SF $z$ axis. The basis set used in Eq.~(\ref{eq:psi}) is similar to the one used in the previous work of Volpi and Bohn \cite{Volpi:02} and Krems and Dalgarno \cite{Krems:04} for open-shell $^2\Sigma$ and $^3\Sigma$ molecules colliding with structureless atoms. The only difference is that our basis functions $|IM_I\rangle$ describe the nuclear spin degrees of freedom in $^1\Sigma$ molecules rather than the electron spins of $^2\Sigma$ and $^3\Sigma$ molecules.

The projection of the total angular momentum  $M=M_N+M_I+M_L$, unlike the total angular momentum itself, is conserved for collisions in a magnetic field. 
Substituting Eq.~(\ref{eq:psi}) into the time-independent Schr$\ddot{o}$dinger equation, $\hat{H}|\Psi\rangle=E|\Psi\rangle$, where $E$ is the total energy, we obtain a system of coupled-channel (CC) equations for the expansion coefficients $F^M_{NM_N IM_I LM_L}$ (omitting the initial quantum numbers $N_i$, $M_{N_i}$, $I_i$, $M_{I_i}$, $L_i$, $M_{L_i}$ for simplicity)
\begin{equation}
\begin{aligned}
\left[ \dfrac{d^2}{dR^2}+2\mu E-\dfrac{L(L+1)}{R^2} \right]F^M_{NM_NIM_ILM_L}(R)=\\
2\mu \sum_{N',M_N',M_I'}\sum_{L'M_L'}\langle NM_NIM_ILM_L|\hat{V}+ \hat{H}_{\rm mol}|N'M_N'I'M_I'L'M_L'\rangle F^M_{N'M_N'IM_I'L'M_L'}(R)
\end{aligned}
\label{eq:cc}
\end{equation}
where the summation is carried out over all the channels included in the basis set. The CC equations (\ref{eq:cc}) are parametrized by the matrix elements of the molecular Hamiltonian and of the interaction potential in the direct-product basis (\ref{eq:psi}). Below we describe the evaluation of these matrix elements.

We begin with the matrix elements of the Hamiltonian of an isolated $^1\Sigma$ molecule  (\ref{Hmol}).
Because the rotational Hamiltonian  is independent of the nuclear spin and orbital degrees of freedom, it is diagonal in $M_I$, $L$, and $M_L$:
\begin{equation}
\begin{aligned}
&\langle NM_NIM_ILM_L|\hat{H}_{\rm rot}|N'M_N'IM_I'L'M_L'\rangle=\delta_{N,N'}\delta_{M_N,M_N'}\\
& \times \delta_{M_I,M_I'}\delta_{L,L'}\delta_{M_L,M_L'}{[B_eN(N+1)-D_vN^2(N+1)^2]}
\end{aligned}
\label{eq:rot}
\end{equation}
The matrix elements of the hyperfine Hamiltonian $\hat{H}_\text{hf}$ are obtained by  expanding the spin-rotation interaction $A\hat{\mathbf{I}}\cdot \hat{\mathbf{N}}$ in terms of the raising and lowering operators $\hat{I}_\pm$ and $\hat{N}_\pm$ \cite{Zare:88}
\begin{equation}
\begin{aligned}
\langle NM_NIM_ILM_L|\hat{H}_{\rm hf}|N' M_N' I' M_I' L' M_L'\rangle=
\delta_{L,L'}\delta_{M_L,M_L'} A [\delta_{M_N,M_N'}\delta_{M_I,M_I'}M_N'M_I' \\ +
\dfrac{1}{2}(C_+^{I,M_I'}C_-^{N,M_N'}\delta_{M_I,M_I'+1}\delta_{M_N,M_N'+1}+
C_-^{I,M_I'}C_
+^{N,M_N'}\delta_{M_I,M_I'-1}\delta_{M_N,M_N'+1})],
\end{aligned}
\label{eq:hf}
\end{equation}
where $C_{\pm}^{j,m} =\sqrt{j(j+1)-m(m\pm 1)}$.
The matrix elements of the Zeeman interaction are diagonal in the uncoupled basis since the basis states $|IM_I\rangle$ are eigenstates of $\hat{I}^2$ and $\hat{I}_z$
\begin{equation}
\begin{aligned}
\langle NM_NIM_ILM_L|\hat{H}_{\rm Z} |N'M_N'IM_I'L'M_L'\rangle=\\
\delta_{N,N'}\delta_{M_N,M_N'}
\delta_{L,L'}\delta_{M_L,M_L'} \delta_{M_I,M_I'}  (-g_N\mu_N M_N B 
-g_I\mu_N M_I B) 
\\ -\delta_{N,N'}\delta_{M_N,M_N'}
\delta_{L,L'}\delta_{M_L,M_L'} \delta_{M_I,M_I'}
B^2 \dfrac{3M_N^2-N(N+1)}{3(2N-1)(2N+3)}(\chi_{\parallel}-\chi_{\perp})
\end{aligned}
\label{eq:odz}
\end{equation}
where the  matrix elements of the diamagnetic Zeeman interaction are proportional to the difference between $\chi_{\parallel}$ and $\chi_\perp$, the parallel and perpendicular components of the diamagnetic susceptibility tensor of CO  (see Eq.~(8.140) of Ref.~\citenum{Brown:03}).
Test calculations show that the diamagnetic Zeeman interaction becomes  noticeable only at high magnetic fields ($B>1$~T). 

The atom-molecule interaction potential is rotationally invariant and independent of the nuclear spin. Hence, its matrix elements are diagonal in the total angular momentum projection $M$ and in the nuclear spin projection $M_I$  \cite{Krems:04}
\begin{equation}
\begin{aligned}
 \langle N M_{N} IM_I L M_L| 
V_{\lambda}(R,r,\theta)| 
N' M'_NI' M'_IL' M'_L\rangle
=\delta_{M_{I}, M'_{I}}(-1)^{M'_L-M_N} \\
\times
\big[ (2L+1)(2L'+1) \big]^\frac{1}{2}\big[(2N+1)(2N'+1) \big]^\frac{1}{2}\sum_{\lambda}  V_{ \lambda}(R)
 \begin{pmatrix}
L & \lambda & L'\\
0 & 0 & 0
\end{pmatrix}\\
\times
\begin{pmatrix}
L & \lambda & L'\\
-M_L & M_L-M'_L & M_L'
\end{pmatrix} 
 \begin{pmatrix}
N & \lambda & N'\\
0 & 0 & 0
\end{pmatrix}
\begin{pmatrix}
N & \lambda & N'\\
-M_{N} & M_{N}-M_{N'} & M_{N'}
\end{pmatrix},
\end{aligned}
\label{eq:v}
\end{equation}
where the Legendre coefficients $V_\lambda(R)$ are obtained by expanding the interaction potential energy surface (PES) in Legendre polynomials $V(R,\theta)=\sum_\lambda V_\lambda(R)P_\lambda(\cos\theta)$ (see Section III A for details).

To obtain the full state-to-state reactance ($K$) and scattering ($S$) matrices, we match the asymptotic solutions of CC equations (\ref{eq:cc}) to the standard asymptotic form given by linear combinations of the Riccati-Bessel and Neumann functions at large $R$ \cite{Johnson:73}.

The state-to-state scattering cross sections are related to the  $S$ matrix elements at a given collision energy $E$
\begin{equation}
\begin{aligned}
 \sigma_{\gamma\to\gamma'}(E)= \dfrac{\pi}{k^2_\gamma}\sum_M\sum_{LM_L}\sum_{L'M_L'}
  |\delta_{LM_L,L'M_L'}\delta_{\gamma \gamma'}-S_{\gamma LM_L,\gamma'L'M_L'}^{M}|^2,
\end{aligned}
\label{eq:xsection}
\end{equation}
where $\gamma$ and $\gamma'$ refer to the eigenstates of the isolated molecule's Hamiltonian (\ref{Hmol}) in the presence of a magnetic field, $|\gamma\rangle=\sum_{NM_NM_I}C_{\gamma,NM_NM_I}(B)|NM_N\rangle |IM_I\rangle$, and $k_\gamma=\sqrt{2\mu E}$ is the collision wavevector. The matrix of solutions of  CC equations is transformed to the eigenstate basis before the application of scattering boundary conditions \cite{Krems:04}.

The thermal state-to-state rate coefficients at temperature $T$ are obtained by averaging the cross sections over the Maxwell-Boltzmann velocity distribution  \cite{Yang2005jcp} 
\begin{equation}
\begin{aligned}
K_{\gamma\to\gamma'}(T)=\left(\dfrac{8}{\pi\mu k_B^3T^3}\right)^{\frac{1}{2}} \int_{0}^{\infty} \sigma_{\gamma\to\gamma'}(E) E {\rm  exp}\left(-\dfrac{E}{k_BT}\right)dE,
\end{aligned}
\label{eq:rate}
\end{equation}
where $k_B$ is the Boltzmann constant and $T$ is the  temperature.

\subsection*{Computational details}

We use the following spectroscopic constants of $^{13}$C$^{16}$O to parametrize the Hamiltonian in Eq.~\eqref{Hmol}
$B_e=55.101 $ GHz \cite{klapper2000sub}, $D_v=1.676 \times10^{-4} $ GHz \cite{klapper2000sub},  $A=3.27 \times10^{-5}$ GHz \cite{meerts1977electric}, $g_N=-0.2595$\cite{meerts1977electric},  $g_I=1.40482$ \cite{EasySpin}, and the diamagnetic susceptibility anisotropy  $(\chi_{\parallel}-\chi_{\perp})=-6.85829 \times10^{-14} \, {\text{cm}^{-1}}/{\text{T}^2}$  \cite{meerts1977electric}.

We use the log-derivative approach \cite{Johnson:73,Manolopoulos:86} to numerically integrate the CC equations (\ref{eq:cc})  for He~+~$^{13}$C$^{16}$O collisions on a radial grid from $R_\text{min}=3.0$ $a_0$ to $R_\text{max}=110.0$ $a_0$ with a constant grid step 0.02 $a_0$. While here we are only interested in transitions between the Zeeman states in the first two rotational manifolds ($N=0$ and $1$), the CC basis must  include closed channels to ensure numerical convergence of the calculated cross sections. In order to ensure the convergence, we found that it is necessary to include 11 lowest rotational states of CO in our calculations. For collision energies below (above) 2 K we include  all partial waves with  $L\le10$ ($L\le13$), which results in a total of 288 (450) coupled channels.

\textcolor{red}{
To calculate  collision rates, we averaged the calculated cross sections according to Eq.~(\ref{eq:rate}) on a grid of collision energies from $1.44 \times 10^{-6}$ to 14.4 K with 102 grid points fitted with cubic splines.
In doing so, we found that calculating the upward excitation rates (e.g., for the $\ket{2}\to \ket{3}$ transition) is challenging because excitation transitions are energetically forbidden at collision energies below the $N=1$ threshold (5.3 K), and their cross sections   increase sharply from zero to a finite value above the threshold. As this behavior is very hard to fit, we use the principle of detailed balance 
\begin{equation}
\begin{aligned}\label{det_bal}
\frac{k_{f\to i}(T)}{k_{i \to f}(T)}= e^{\frac{E_f-E_i}{k_B T}}
\end{aligned}
\end{equation}
to obtain the desired excitation  rates, where $E_i$ and $E_f$ are the energies of the initial and final molecular states involved in the transition.}

We use an accurate {\it ab initio} PES for He-CO developed by Heijmen {\it et al.} \cite{heijmen1997new} using a symmetry-adapted perturbation theory approach \cite{moszynski1995ab}.
This PES was used in several quantum scattering calculations by Balakrishnan {\it et al.} \cite{Balakrishnan2000jcp} and by Wang {\it et al.} \cite{Yang2005jcp}, which focused on rovibrational transitions  in $^{12}$C$^{16}$O induced by ultracold collisions with He atoms. Low-energy scattering resonances in He~+~CO collisions were observed  in a merged beam experiment and compared to theoretical calculations using several PESs \cite{Bergeat:15}.  We compare our results against the previous calculations \cite{Balakrishnan2000jcp,Yang2005jcp} in \textcolor{red}{Appendix B} to test our He~+~CO scattering code. The He-CO PES is weakly anisotropic, with a single  minimum of depth -23.734 cm$^{-1}$ located at $R=6.53$ $a_0$, $\theta=$ 48.4$^\circ$, and $r=2.132\,a_0$ (the equilibrium bond length of CO).  We expand the $\theta$ dependence of the PES in 12 Legendre polynomials.

The previous calculations did not account for the nuclear spin of the $^{13}$C$^{16}$O isotopologue, which is the subject of interest here. Because the original He-CO PES was defined in the Jacobi coordinate system with the origin at the center of mass of $^{12}$C$^{16}$O, we need to rescale the PES to account for the shift of the center of mass. The rescaling procedure is described in \textcolor{red}{Appendix C}.

\section{Results and discussion}

In this section, we apply the theory developed above to study the nuclear spin dynamics in cold collisions of $^{13}$C$^{16}$O molecules with  $^4$He atoms. We will present state-to-state scattering cross sections and collision rates for all initial Zeeman levels of $^{13}$C$^{16}$O ($N=0,1$) in collisions with $^4$He atoms. 
We also calculate the magnetic field dependence of inelastic collision rates and explain it using  a simple model.
Finally, we use the computed collision rates to explore the thermalization dynamics of a single nuclear spin state of $^{13}$C$^{16}$O immersed in a cold buffer gas of $^4$He, and estimate the nuclear spin relaxation times.

\subsection{Energy levels of $^{13}$C$^{16}$O}

\begin{figure}[t]
 \includegraphics[width=10cm ]{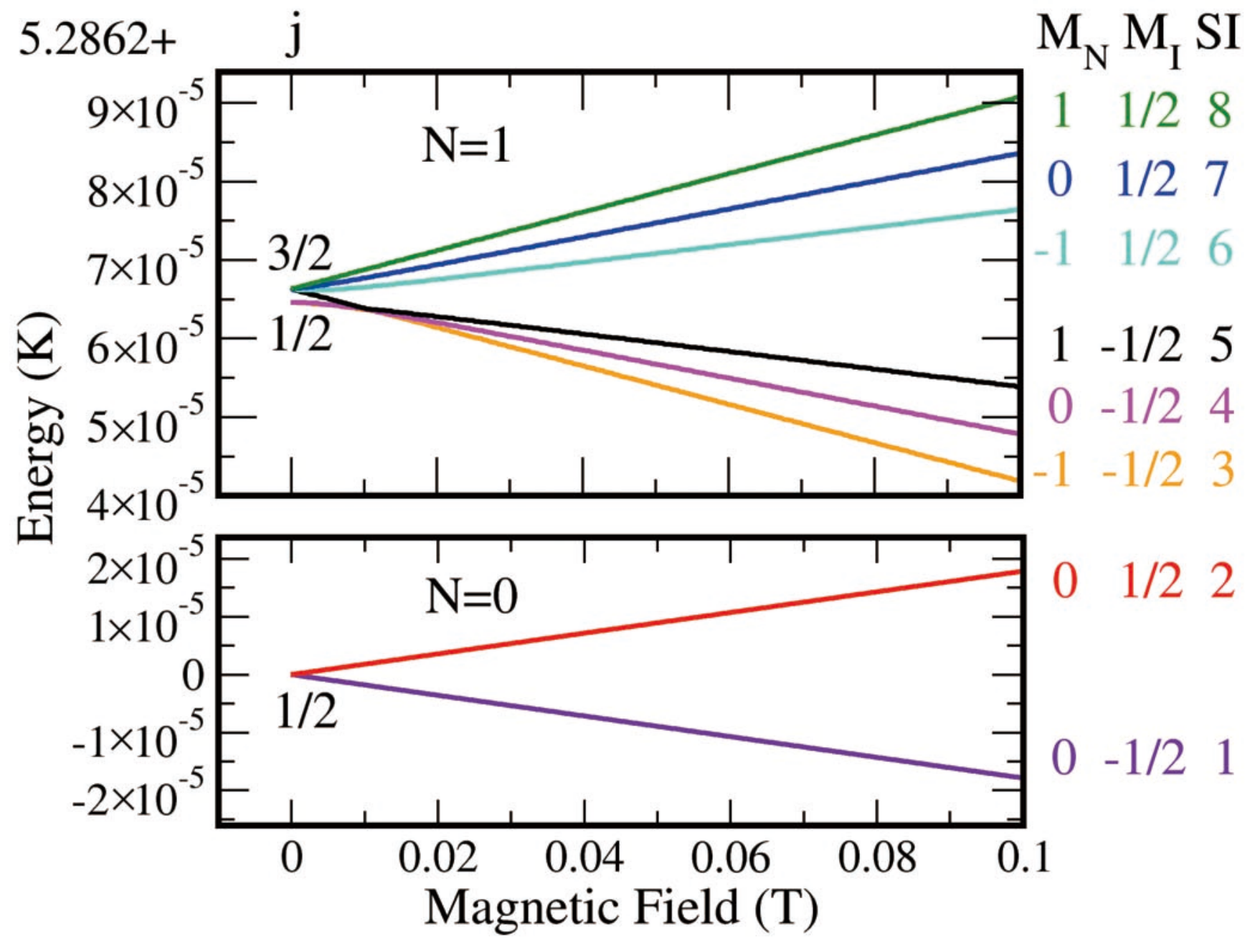}
 \caption{Zeeman energy levels of $^{13}$C$^{16}$O as a function of magnetic field. Each level is labeled by the quantum numbers $j$ (in the low-field limit) and $N,M_N,M_I$ (in the high-field limit) and by \textcolor{red}{its state index (SI). 1~K = 0.695~cm$^{-1}$.} Top and bottom panels show the levels in the rotational manifolds $N=1$ and $N=0$, respectively.}
   \label{fig:elevels}
\end{figure}

Figure \ref{fig:elevels} displays the Zeeman levels of $^{13}$C$^{16}$O as a function of magnetic field. The electronic spin of $^{13}$C$^{16}$O is zero and the energy splittings are determined by an interplay between the hyperfine (spin-rotation) interaction due to the nuclear spin of $^{13}$C ($I=1/2$)  and the Zeeman interaction given by Eq.~(\ref{eq:odz}). Throughout the rest of this paper, we will only consider the $^{13}$C$^{16}$O isotope, and hence will omit the isotope labels.

 The largest splitting of the energy levels is due to the rotational structure, which gives rise to two manifolds of rotational states labeled $N=0$ and $N=1$ separated by $3.6761$~cm$^{-1}$, approximately twice the rotational constant of CO. 
 At zero magnetic field the  nuclear spin-rotation interaction splits all $N\ge 1$ energy levels into doublets with $j=N-1/2$, $N+1/2$, where $j=|\mathbf{N}+\mathbf{I}|$ is the total angular momentum of the molecule. Note that $j$ is a good quantum number at zero field. The splitting between the $j=1/2$ and $j=3/2$ levels in the $N=1$ manifold is $\frac{3}{2}$ $A$ or 2.4 $ \times 10^{-6}$ cm$^{-1}$.

 At high magnetic fields the  Zeeman interaction overcomes the hyperfine interaction, $j$ is no longer conserved, and the energy levels approach the eigenstates of the Zeeman Hamiltonian {[Eqs.~(\ref{eq:odz})]}  $|NM_N\rangle|IM_I\rangle$ with the good quantum numbers $N$, $M_N$, and $M_I$  indicated in Fig.~\ref{fig:elevels}.
  In the $N=0$ manifold, the nuclear spin-rotation interaction vanishes and there are two Zeeman levels with $ M_I=\pm 1/2$ as shown in  the bottom panel of Fig.~\ref{fig:elevels}.
  The $N=1$ manifold contains six Zeeman levels  $|N,M_N,M_I\rangle$ with $N=1$, $M_N=-1,0,+1$, and $M_I=\pm 1/2$. To facilitate the following discussion, we will assign a State Index (SI) to each Zeeman energy level as shown in Fig.~\ref{fig:elevels} in the order of increasing energy. It is worth noting that the states $\ket{3}$, $\ket{4}$ and $\ket{5}$ cross at $B \simeq 0.01$~T, where the nuclear spin-rotation interaction becomes comparable to the Zeeman interaction. The crossings are not avoided because the states have different values of $m$, the projection of $j$ on the magnetic field axis, which is a good quantum number for a molecule in an external magnetic field.

\subsection{Nuclear spin dynamics in cold He-CO collisions: Cross sections and rate constants}

\begin{figure}[t]
 \centering
 \includegraphics[width=0.6\textwidth]{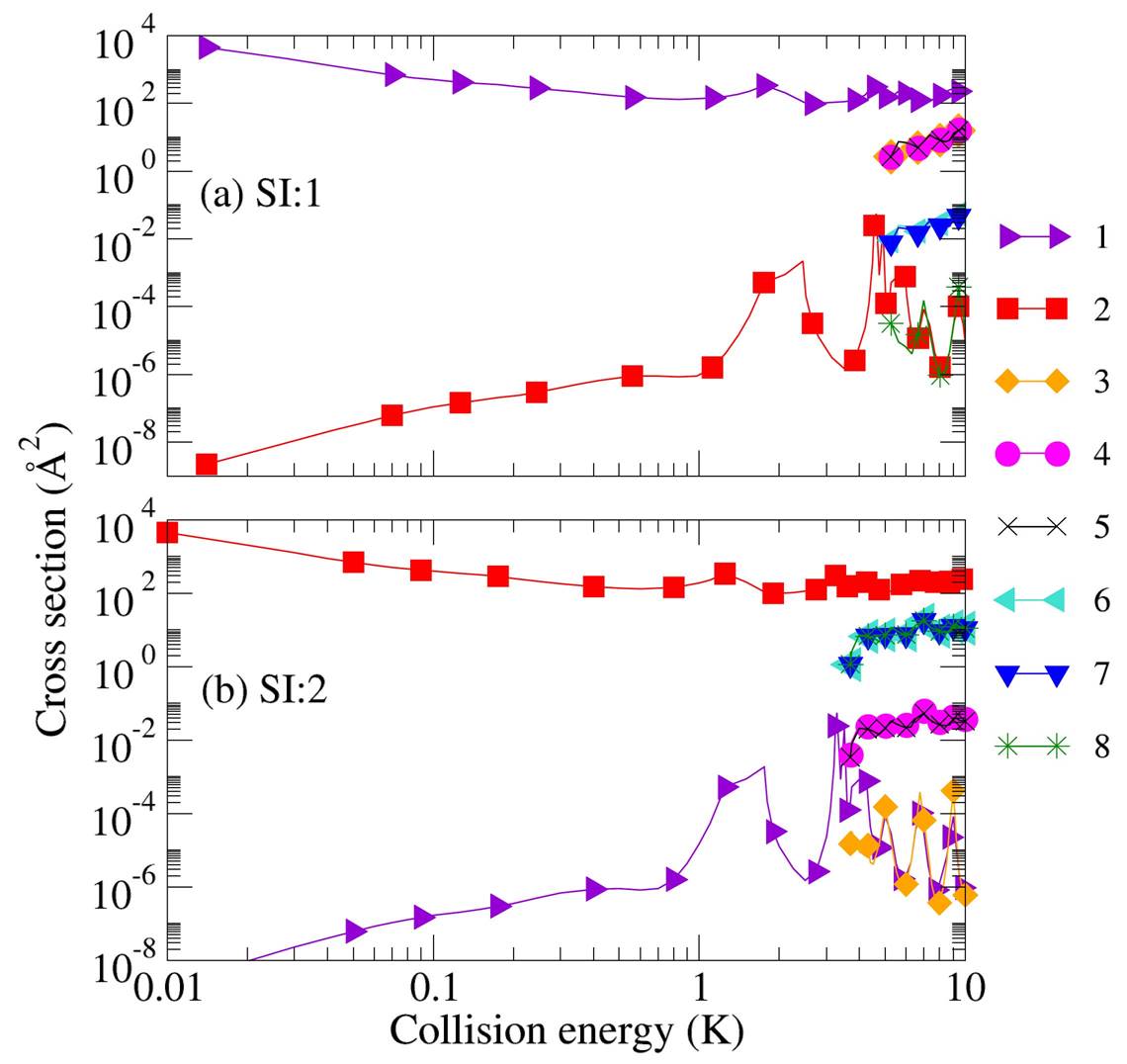}
 \caption{State-to-state cross sections for He~+~CO collisions for the initial Zeeman states in the $N=0$ manifold: $|M_I=-1/2\rangle$ (a) and $|M_I=1/2\rangle$ (b) in a magnetic field of $0.05$~T plotted as a function of collision energy. The cross sections for the transitions $|1\rangle {\to} |2\rangle,|8\rangle$ and $|2\rangle {\to} |1\rangle,|3\rangle$ are multiplied by 10$^6$ for better visibility. \textcolor{red}{Dashed lines indicate the energy gap between the $N=0$ and $N=1$ levels ($\simeq 5.3$~K). The initial states are indicated in each panel. The various color-coded symbols indicate the final state indices (SIs) specified in Fig.~\ref{fig:elevels}.} \textcolor{red}{The symbols do not represent the actual number of calculated points, which is 20 per order of magnitude (i.e., 20 points between 0.01 and 0.1 cm$^{-1}$, etc.)}}
  \label{fig:xsection12}
\end{figure}

We begin by considering the transitions between  the two lowest nuclear spin levels of CO, $|N=0, M_I=\pm 1/2\rangle$ induced by cold collisions with He atoms. Figure~\ref{fig:xsection12} shows the cross sections for transitions out of the ground ($|M_I=-1/2\rangle$) and the first excited ($|M_I=1/2\rangle$) Zeeman levels  labeled as 1 and 2 in  Fig.~\ref{fig:elevels}. 
The elastic cross sections $\sigma_{1\to1}$ and $\sigma_{2\to2}$ are on the order of 100-10$^4$~\AA$^2$ between 10 mK and 10 K, as is typical for cold atom-molecule collisions. The inelastic cross sections for nuclear spin-changing transitions within the $N=0$ manifold $|M_I=1/2 \rangle \leftrightarrow |M_I=-1/2\rangle$ are {\it extremely small}, reaching values below 10$^{-15}$~\AA$^2$ at 10~mK.

Similarly to electron spin relaxation in  collisions of $^2\Sigma$ molecules with atoms \cite{Krems:03,Krems:04b}, nuclear spin relaxation in cold atom-molecule collisions ($E< 2B_e$) proceeds indirectly, via a three-step process. First, the molecule is temporarily  excited to a closed-channel $N=1$ state  due to the coupling with the incident $N=0$ channel  induced by the anisotropy of the interaction potential. In the rotationally excited state, the nuclear spin is flipped by the  nuclear spin-rotation interaction (\ref{eq:hf}). 
Finally, the molecule is de-excited to  the ground $N=0$ state through the interaction anisotropy coupling. Because the nuclear spin-rotation interaction is 2-3 orders of magnitude weaker than the electron spin-rotation interaction in $^2\Sigma$ molecules, nuclear spin relaxation will be suppressed by the factor $\simeq(A/\gamma_{sr})^2$ (i.e., by additional 4-6 orders of magnitude) compared to the electron spin relaxation mediated by the electron spin-rotation interaction $\gamma_{sr}\mathbf{N}\cdot \mathbf{S}$. \cite{Krems:03,Krems:04}

At collision energies  exceeding the rotational spacing ($E\geq 2B_e \simeq 5.3$~ K for CO), $N=0\to 1$   rotational  excitation transitions become energetically allowed, and the spin relaxation cross sections $\sigma_{2\to1}$ and $\sigma_{1\to2}$ increase dramatically. As in the case of electron spin relaxation \cite{Krems:03,Krems:04,Maussang:05,Singh:13},  this occurs because the $N=0\to 1$ transitions can now  populate the open $N=1$ states directly (rather than transiently).

The cross sections for nuclear spin-conserving rotational excitation increase slightly with collision energy  as shown in Fig.~\ref{fig:xsection12} and are the largest between the spin-down energy levels $|1\rangle$ and $|3{-}5\rangle$ or between the spin-up energy levels  $|2\rangle$ and $|6{-}8\rangle$.  On the other hand, the nuclear spin-flipping rotational excitation transitions    $|1\rangle \to |6{-}8\rangle$, and $|2\rangle \to |3{-}5\rangle$ are suppressed by  a factor of $\simeq$100. This is an example of a nuclear spin selection rule discussed in more detail below.
We observe several Feshbach resonances in the collision energy dependence of the cross sections $\sigma_{1\to2}$, $\sigma_{1\to 8}$,  $\sigma_{2\to1}$, and $\sigma_{2\to3}$ above 1 K.

\begin{figure}[t]
 \centering
 \includegraphics[width=0.75\textwidth]{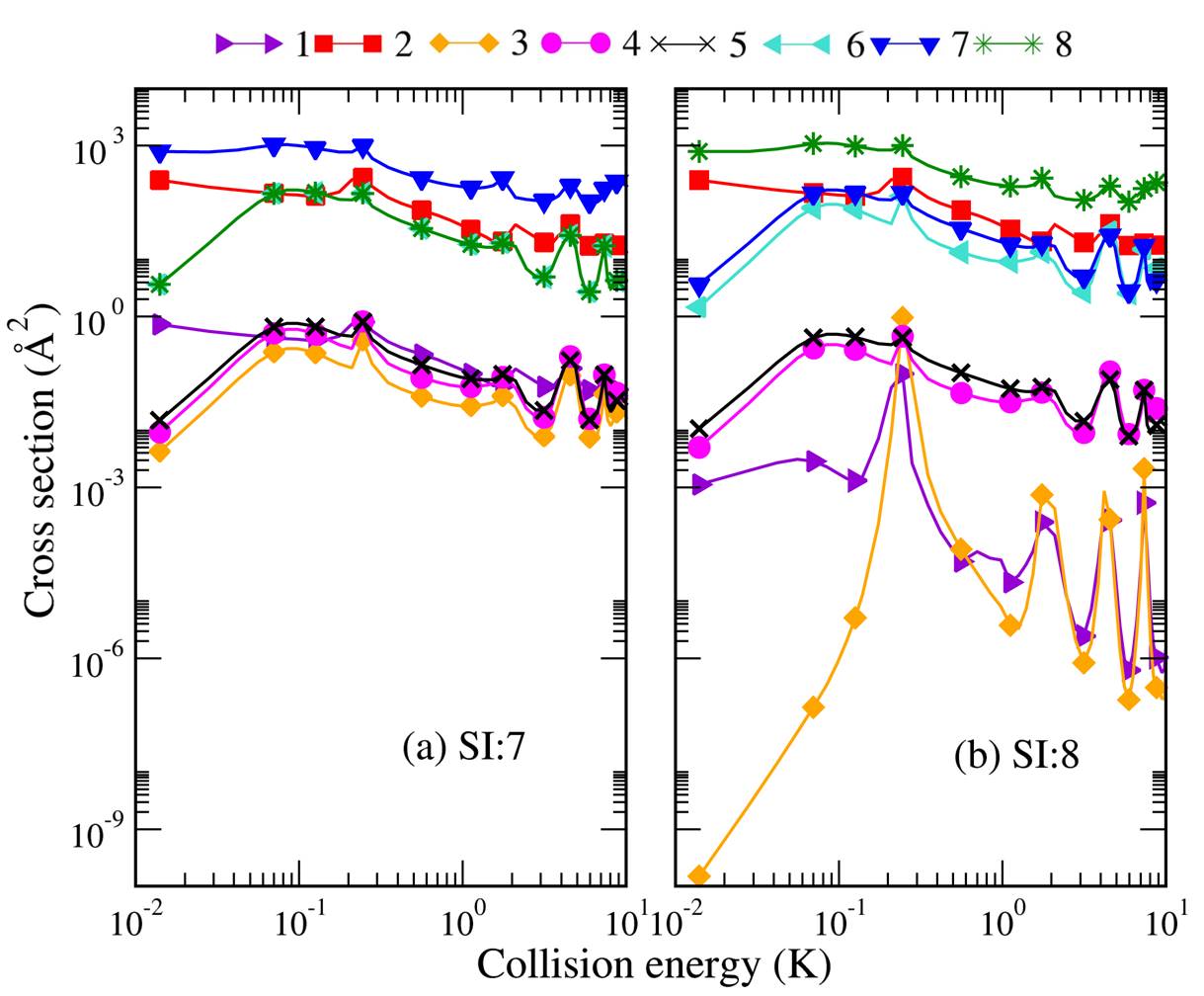}
 \caption{State-to-state cross sections of He~+~CO collisions for the initial Zeeman states $\ket{7}$ and $\ket{8}$ in the $N=1$ manifold plotted as a function of collision energy.  The initial states are indicated in each panel. \textcolor{red}{The various color-coded symbols indicate the final state indices (SIs) specified in Fig.~\ref{fig:elevels}. The cross sections for the transitions $|8\rangle {\to} |1\rangle $ and $|8\rangle {\to} |3\rangle$ are multiplied by 10$^6$ for better visibility.}  \textcolor{red}{The symbols do not represent the actual number of calculated points, which is 20 per order of magnitude (i.e., 20 points between 0.01 and 0.1 cm$^{-1}$, etc.)}}
  \label{fig:xsection78}
\end{figure}

\begin{figure}[t]
 \centering
 \includegraphics[width=0.75\textwidth]{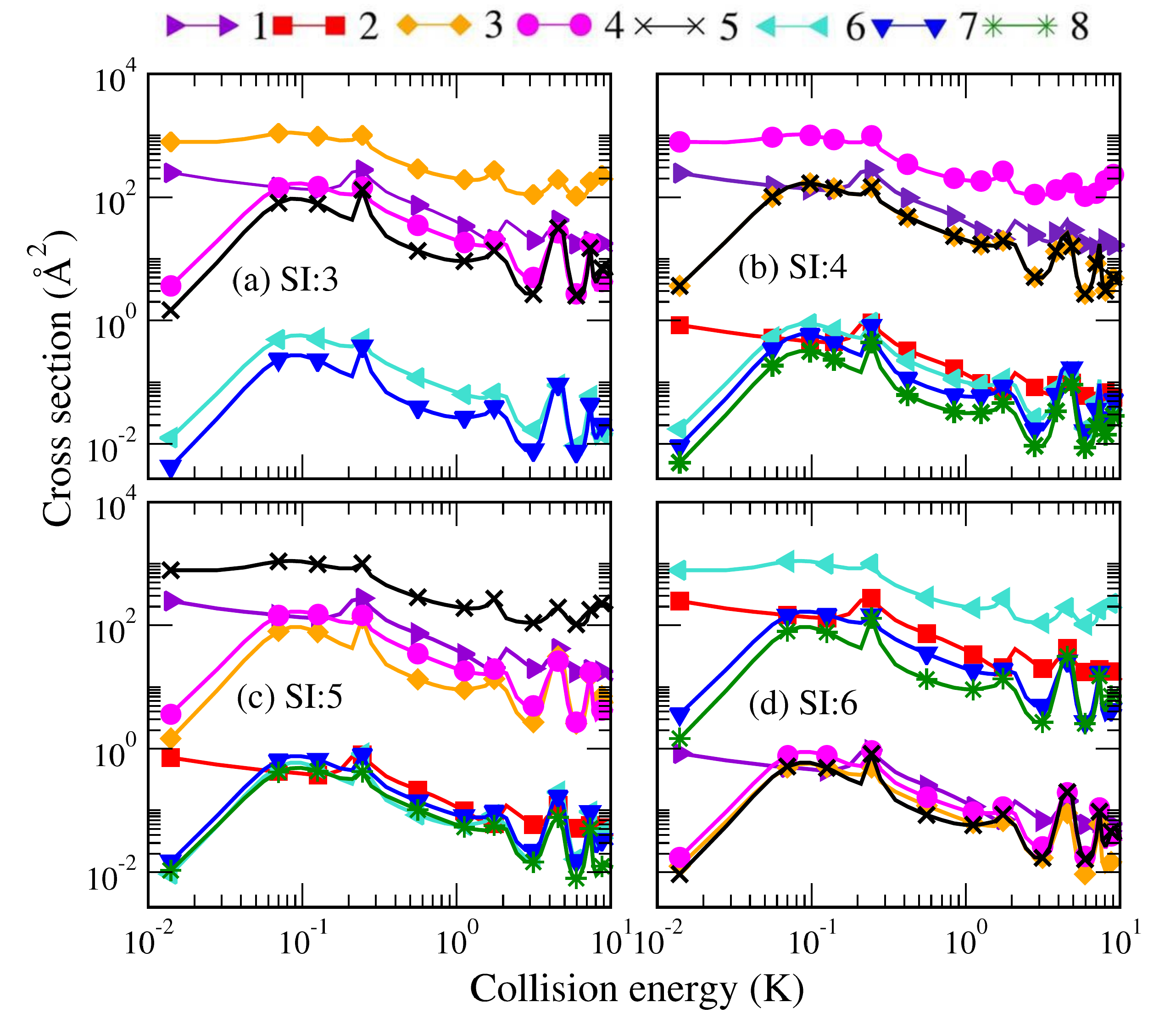}
 \caption{State-to-state cross sections of He~+~CO collisions for the initial Zeeman states $\ket{3}{-}\ket{6}$ in the $N=1$ manifold plotted as a function of collision energy.  The initial states are indicated in each panel. \textcolor{red}{The various color-coded symbols indicate the final state indices (SIs) specified in Fig.~\ref{fig:elevels}. The cross sections for   group-III transitions  $|3\rangle {\to} |2\rangle$ and $|3\rangle {\to} |8\rangle$ are negligibly small and hence not shown in panel (a).} \textcolor{red}{The symbols do not represent the actual number of calculated points, which is 20 per order of magnitude (i.e., 20 points between 0.01 and 0.1 cm$^{-1}$, etc.)} }
  \label{fig:xsection3to6}
\end{figure}

\textcolor{red}{In Figs.~\ref{fig:xsection78} and \ref{fig:xsection3to6}}, we show the cross sections for elastic scattering and rotational relaxation in He~+~CO collisions with CO molecules initially prepared in the different Zeeman sublevels of the $N=1$ excited rotational state. For the fully spin-polarized initial states $|8\rangle$ and $|3\rangle$,  we observe three groups of transitions \textcolor{red}{as shown in Fig.~\ref{fig:xsection78}}. The dominant relaxation channels are $|8\rangle \to |2\rangle$ ($|1, 1, 1/2 \rangle \to |0, 0, 1/2\rangle$), $|8\rangle \to |6\rangle$ ($|1, 1, 1/2 \rangle \to |1, -1, 1/2\rangle$)and $|8\rangle \to |7\rangle$ ($|1, 1, 1/2 \rangle \to |1, 0, 1/2\rangle$) (group I). All of these transitions conserve the nuclear spin projection $M_I$ but change either $N$ or $M_N$.
The second most probable are the nuclear spin-flipping transitions $|8\rangle \to |4\rangle$ ($|1, 1, 1/2 \rangle \to |1, 0, -1/2\rangle$) and $|8\rangle \to |5\rangle$ ($|1, 1, 1/2 \rangle \to |1, 1, -1/2\rangle$) (group II), which change $M_I$ but conserve $N$. The cross sections in group II are $\simeq10^3$ times smaller than those in group I.
In turn, the least probable transitions from group III,  $|8\rangle \to |1\rangle$ ($|1, 1, 1/2 \rangle \to |0, 0, -1/2\rangle$) and $|8\rangle \to |3\rangle$ ($|1, 1, 1/2 \rangle \to |1, -1, -1/2\rangle$), are further suppressed by a factor of $\simeq 10^3$ compared to group II transitions.  The final states in group III are also fully spin-polarized, with a maximal difference between the initial and final values of $M_N$ and $M_I$. This suppression  is expected because the fully spin-polarized states are not directly coupled by the nuclear spin-rotation interaction, and thus have no direct matrix elements  connecting them. The cross section for the $|8\rangle \to |3\rangle$ transition is extremely small at all collision energies as shown in Fig.~\ref{fig:xsection78}(b).
Similar trends are observed for the  fully spin-polarized initial state $\ket{3}$  
as well as in our recent calculations on electron spin relaxation of rotationally excited CaH molecules in cold collisions with He atoms in a magnetic field \cite{Koyu:22}.


\textcolor{red}{For non fully spin-polarized $N=1$ initial states $\ket{4}-\ket{6}$, and $\ket{7}$ (see Figs.~\ref{fig:xsection3to6}  and Fig.~\ref{fig:xsection78}}) we observe two distinct groups of transitions, one of which (group I) has much larger cross sections than the other (group II). Group-I transitions conserve $M_I$ but change either $M_N$ or $N$ as in the fully polarized case considered above. Group-II transitions are nuclear spin-flipping transitions, which change $M_I$ and/or $M_N$.  There are no strongly suppressed group-III transitions observed above since  the initial  states are not fully spin-polarized. We note that all nuclear spin-flipping transitions (group-II and group-III)  are forbidden in the first Born approximation due to zero nuclear spin overlap between the initial and final states. These transitions occur due to higher-order  couplings through intermediate rotationally excited states mediated by the anisotropy of the interaction potential (see above) \cite{Krems:03,Krems:04}.

The energy dependence of the cross sections shown \textcolor{red}{in Figs.~\ref{fig:xsection78} and \ref{fig:xsection3to6}} displays a number of resonances. Four substantial resonances occur in the cross sections for group-III transitions,  such as $\ket{8}\to \ket{3}, \ket{1}$ and $\ket{3}\to \ket{8}, \ket{2}$ at 0.25 K, 2 K, 4 K and 7.2 K, respectively. We have verified that these resonances are Feshbach resonances, rather than shape resonances, by calculating the partial wave decomposition of the cross sections \eqref{eq:xsection}, and finding that no single partial wave dominates at the resonance energies. \textcolor{red}{The likely reason why Feshbach resonances are absent in the Group I and, to a less extent, Group II  transitions is the large magnitude of the  cross sections for these transitions, leading to an enhanced decay width of the resonances. As a result, the resonances peaks become suppressed \cite{hutson2007feshbach}.}

\begin{figure}[t]
 \centering
 \includegraphics[width=9 cm]{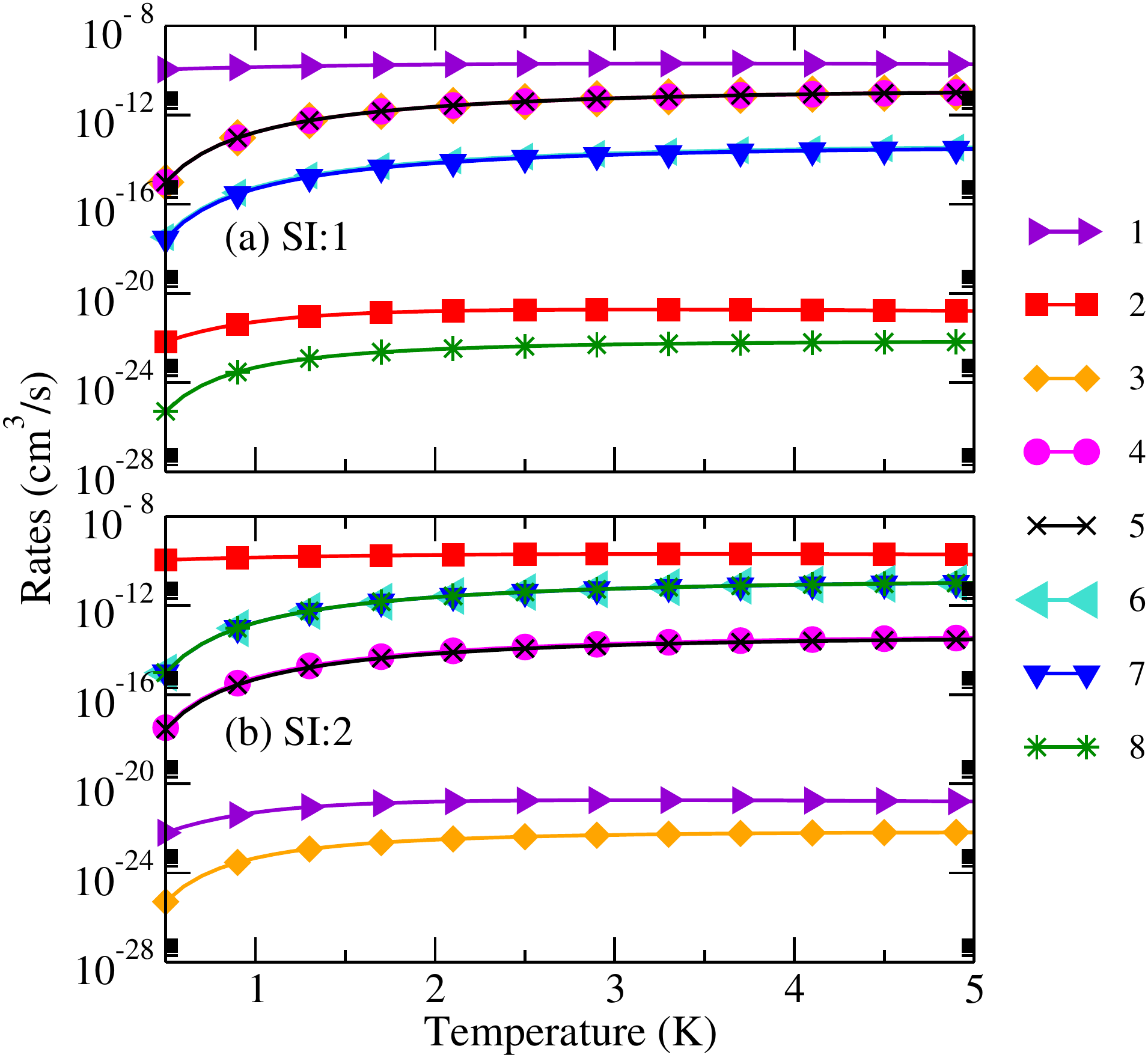}
 \caption{Thermally averaged rate coefficients for He~+~CO collisions plotted as a function of temperature for the initial hyperfine-Zeeman states $\ket{1}$ and $\ket{2}$ of CO. The magnetic field is 0.05~T. \textcolor{red}{The initial states are indicated in each panel. The various color-coded symbols indicate the final state indices (SIs) specified in Fig.~\ref{fig:elevels}.}}
  \label{fig:rate12}
\end{figure}

Typical temperatures used in $^4$He buffer gas cooling experiments range from $\simeq$1~K  to 4 K and higher,  so it is instructive to consider the thermally averaged collision rates. The temperature dependence of the elastic and inelastic collision rates is plotted in Fig.~\ref{fig:rate12}.
The resonances shown in Fig.~\ref{fig:xsection12}   are narrow compared to the width of the Maxwell-Boltzmann energy distribution. 
As a result, the resonances are washed away by thermal averaging and the calculated rate coefficients increase monotonously as a function of temperature between 1~K and 5~K for the transitions of $|1\rangle \to |1\rangle, |2\rangle$ and $|2\rangle \to |1\rangle, |2\rangle$. We observe that the rate coefficients for the excitation transitions $\ket{1},\ket{2} \to \ket{3} - \ket{8}$ shown in the lower panel of Fig.~\ref{fig:rate12}  decay exponentially at lower temperatures and remain nearly  constant at higher temperatures. This is because the rotationally excited final states  become energetically closed at collision energies below 5.3~K, and their cross sections vanish as shown in Fig.~\ref{fig:xsection12}. 

\begin{figure}[t]
 \centering
 \includegraphics[width=13 cm]{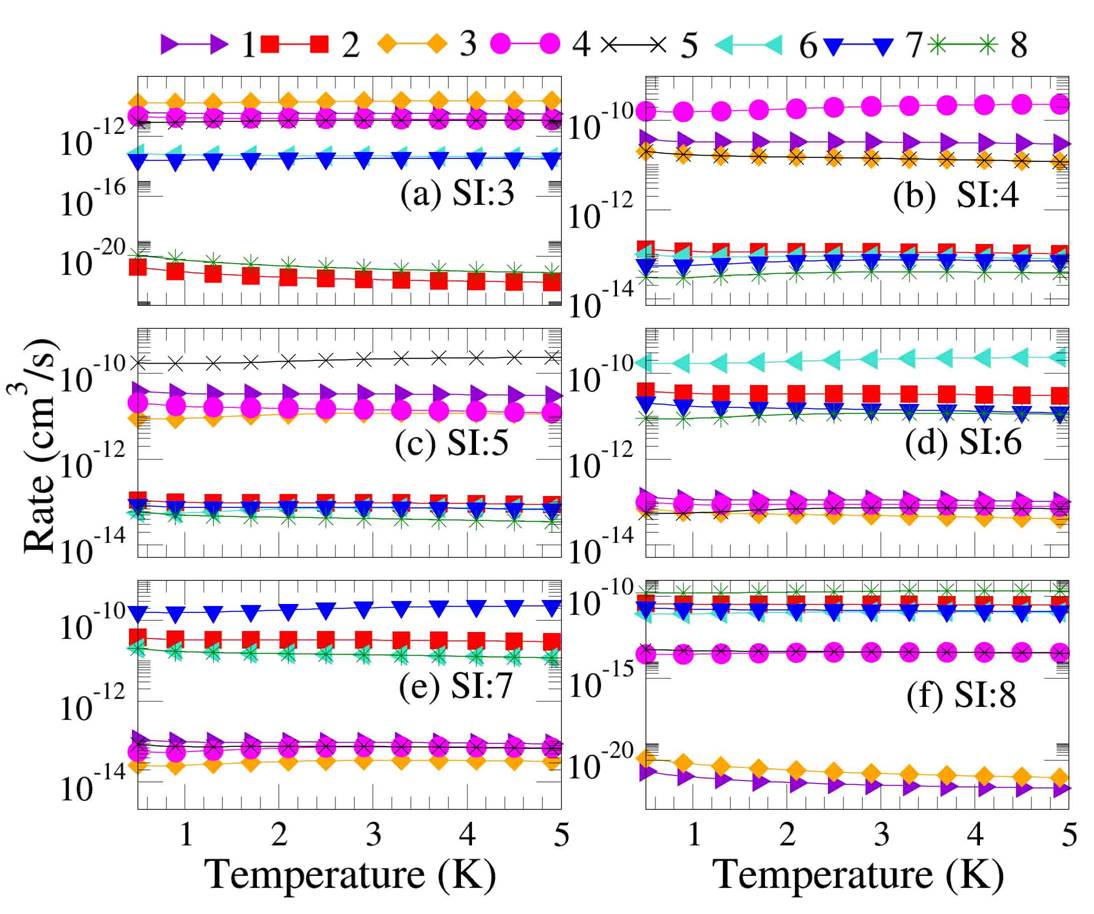}
 \caption{Thermally averaged rate coefficients for He~+~CO collisions for the  rotationally excited initial states $\ket{3}{-}\ket{8}$ in a magnetic field of 0.05~T. \textcolor{red}{The initial states are indicated in each panel. The various color-coded symbols indicate the final state indices (SIs) specified in Fig.~\ref{fig:elevels}.}}
  \label{fig:rate}
\end{figure}

Nevertheless, it should be noted that the {\it rates for rotational excitation via the $\ket{2}\to \ket{3}{-}\ket{8}$ transitions are larger than those for the  $\ket{2}\to \ket{1}$ transition by several orders of magnitude even at  $T=0.5$}~K. 
This is due to the large magnitude of the excitation cross sections at the high-energy tail of the Maxwell-Boltzmann distribution  above 5.3~K (see Fig.~\ref{fig:xsection12}), which  makes a significant contribution to the thermal collision rates even though transitions to rotationally excited closed channels are energetically forbidden below $5.3$~K. 
As shown below, this causes a steep temperature dependence of nuclear spin relaxation times of  CO molecules immersed in a cold buffer gas of He.


The thermally averaged rate coefficients for the transitions out of the initial states in the $N=1$ manifold are shown in Fig.~\ref{fig:rate}. As before, we observe that the rate coefficients are smooth functions of temperature, with the resonant structure present in the energy-dependent cross sections  washed away by thermal averaging.  While all of the nuclear spin-conserving (group-I) transitions have similar transition rates on the order of  10$^{-11}$ cm$^3/$s, the rates of nuclear spin-flipping transitions (group II) are three orders of magnitude smaller.

The rates of group-III transitions involving the fully spin-polarized initial and final states $\ket{8}$, $\ket{3}$, $\ket{2}$, and $\ket{1}$ are further suppressed compared to group-II transitions by several orders of magnitude, reflecting the trend discussed above for the cross sections.  As shown in Figs.~\ref{fig:rate}(a) and \ref{fig:rate}(f) the transition $\ket{8}\leftrightarrow \ket{3}$ is particularly strongly suppressed. This is because the initial and final states are fully spin-polarized and hence they are not coupled by the spin-rotation interaction, just like the $\ket{1}$ and $\ket{2}$ states in the $N=0$ manifold. {\it This remarkable suppression of collisional transitions  is similar to the electron and nuclear spin selection rules in spectroscopy   $\Delta S=0$ and $\Delta I=0$ \cite{JacobsBook}).}  A similar suppression was observed for the transitions between the fully electron spin-polarized $N=1$ Zeeman states of CaH($^2\Sigma^+$) molecules in cold collisions with He atoms \cite{Koyu:22}.

\begin{figure}[t]
 \centering
 \includegraphics[width=11 cm]{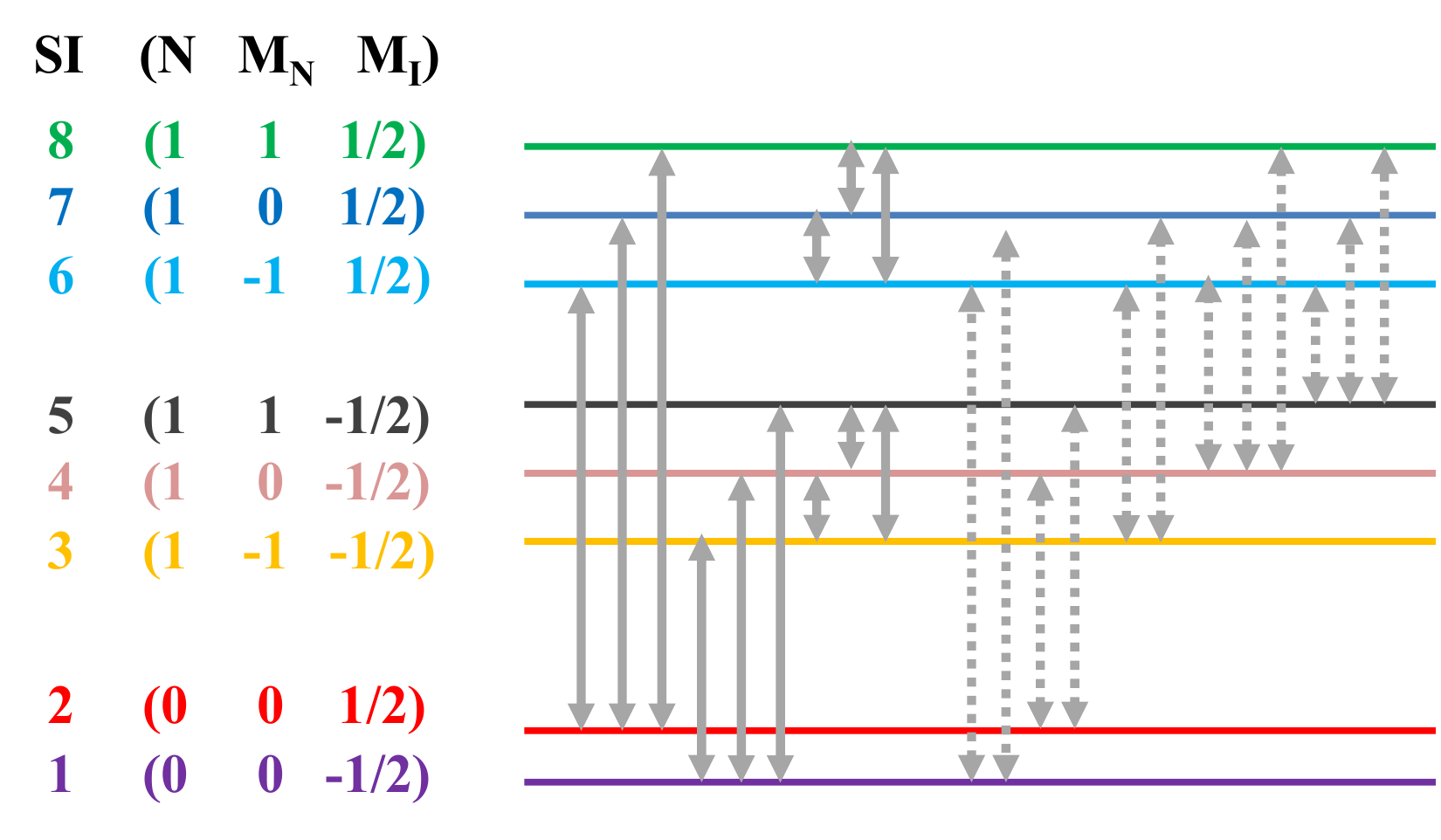}
 \caption{Schematic diagram of relaxation and excitation transitions among the 8 lowest Zeeman levels of CO in cold  collisions with He atoms. The most likely  nuclear spin-conserving (group-I) transitions are marked by solid arrows. Nuclear spin-flipping (group-II)  transitions are marked with dashed arrows. Group-III transitions are forbidden and hence not shown.}
 \label{fig:arrows}
\end{figure}

The dominant relaxation and excitation pathways between the Zeeman sublevels of CO$(N=0,1)$ in cold collisions with He are summarized in Fig.~\ref{fig:arrows}. The most prominent group-I transitions, indicated by solid arrows, conserve the nuclear spin projection.
These transitions occur between spin-up Zeeman levels, ($\ket{2}\leftrightarrow \ket{6{-}8}$, $\ket{6}\leftrightarrow \ket{7}$, $\ket{7}\leftrightarrow \ket{8}$) and $\ket{6}\leftrightarrow \ket{8}$ or between spin-down Zeeman levels ($\ket{1}\leftrightarrow \ket{3{-}5}$, $\ket{3}\leftrightarrow \ket{4}$, $\ket{4}\leftrightarrow \ket{5}$ and $\ket{3}\leftrightarrow \ket{5}$).

The next most efficient are the nuclear spin-flipping transitions from group II marked by dashed arrows in Fig.~\ref{fig:arrows}. Such transitions include  $\ket{1}\leftrightarrow \ket{6{-}7}$, $\ket{2}\leftrightarrow \ket{4{-}5}$, $\ket{3}\leftrightarrow \ket{6{-}7}$ and $\ket{4{-}5}\leftrightarrow \ket{{6}{-}7}$. The group-III spin-flipping transitions involving the nuclear spin-stretched states $\ket{1}$, $\ket{2}$, $\ket{3}$, $\ket{8}$ have extremely small rate coefficients. These forbidden   transitions are  not shown in Fig.~\ref{fig:arrows}.


\subsection{Magnetic field dependence}\label{C}

\begin{figure}[t]
 \centering
 \includegraphics[width=11.0 cm ]{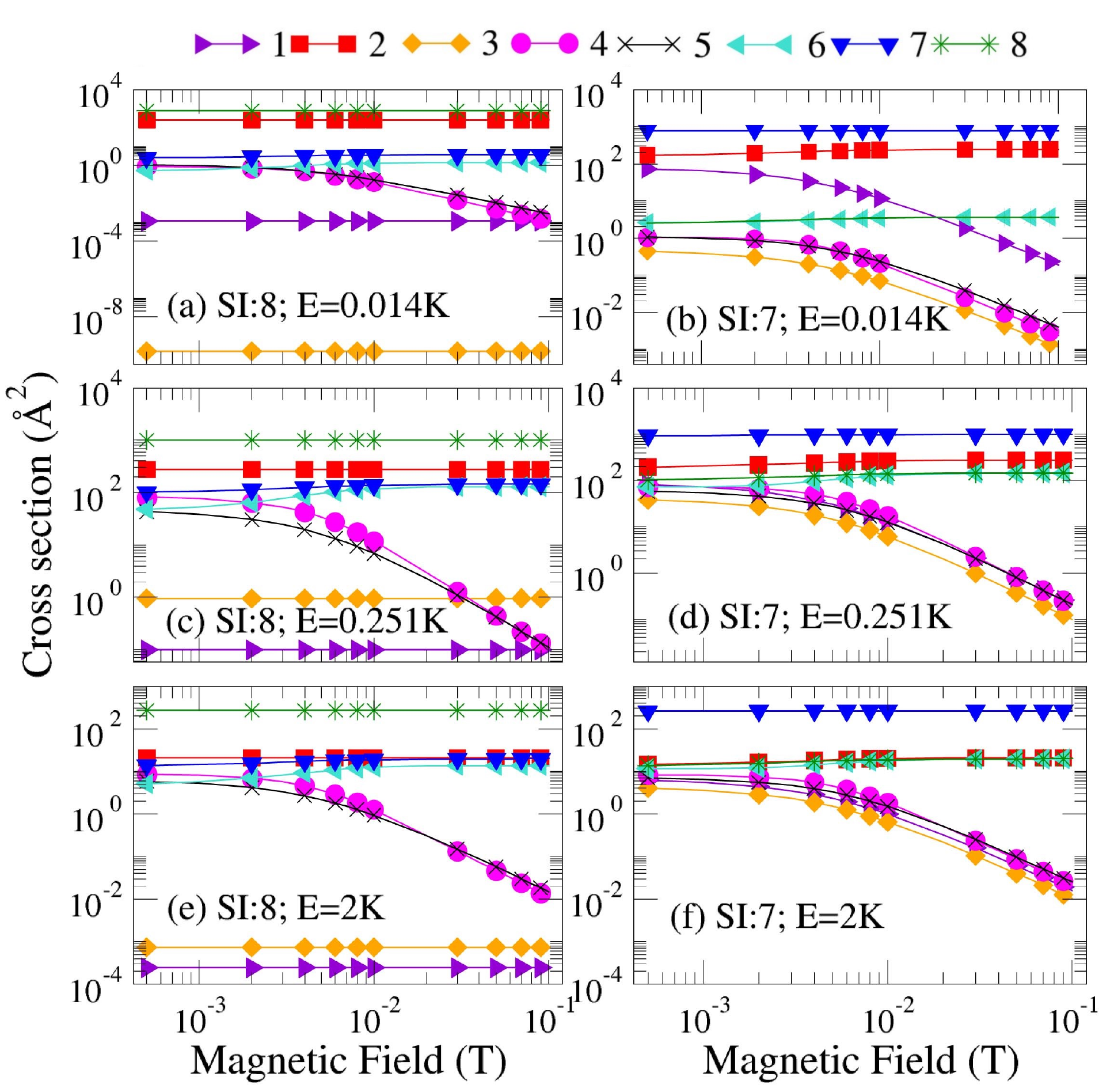}
 \caption{Magnetic field dependence of state-to-state cross sections for  $^{13}$C$^{16}$O$-^4$He collisions. The collision energy and the initial state index  are indicated in each panel. The  cross sections $\sigma_{8\to 1}$ and $\sigma_{8\to 3}$ are multiplied by $10^6$ to aid visibility. \textcolor{red}{The various color-coded symbols indicate the final state indices (SIs) specified in Fig.~\ref{fig:elevels}.}}
 \label{fig:M78}
\end{figure}

In this section, we consider the magnetic field dependence of the state-to-state He~+~CO cross sections. This dependence is depicted in Fig.~\ref{fig:M78} (see also \textcolor{red}{Appendix D}) at $E=0.014$, 0.251, and 2 K and the initial states $\ket{7}$ and $\ket{8}$. 
We note that at these energies, the cross sections for transitions between the $N=0$ Zeeman states $|1\rangle$ and $|2\rangle$ (not shown in Fig.~\ref{fig:M78}) are field-independent.

For the fully spin-stretched initial state $\ket{8}$ we observe three groups of transitions as shown in panels (a), (c), and (e) of Fig.~\ref{fig:M78}, with group III transitions to the fully spin polarized states $\ket{3}$ and $\ket{1}$ having extremely small cross sections, which are independent  of magnetic field (see above).
Transitions to the final states $\ket{2}$, $\ket{7}$, and $\ket{6}$ belong to group I.  These transitions have large cross sections and display no dependence on magnetic field because both the initial and final states  have the same nuclear spin projections. As a result, these transitions are similar to purely rotational transitions, which conserve the nuclear spin.

Importantly, group II transitions
$\ket{8}\to \ket{4}$ and $\ket{8}\to \ket{5}$ show a pronounced magnetic field dependence. These transitions change the nuclear spin projection, and the final states  are superpositions of basis states $\ket{NM_NM_I}$ with different $M_N$ and $M_I$ mixed by the nuclear spin-rotation interaction. 

In Figs.~\ref{fig:M78}(b), (d), and (f) we observe similar trends in the magnetic field dependence of the transitions originating from the initial state $\ket{7}$, except that there is no group-III transitions as state $\ket{7}$ is not fully spin-polarized. As many as 4  transitions out of this state display a marked magnetic field dependence, and  the final states  populated by these transitions ($\ket{1}$, $\ket{3}$, $\ket{4}$, and $\ket{5}$) have the opposite  nuclear spin projection to  the initial state state $\ket{7}$.

   The magnetic field dependence can be explained by examining the initial and final Zeeman states involved in these group-II  transitions. 
 At any finite magnetic field and $N>0$, $M_N $ and $M_I$ are not good quantum numbers because of the nuclear spin-rotation interaction mixing. 
 The molecular eigenstates $\ket{4}$, $\ket{5}$, $\ket{6}$, and $\ket{7}$ can be written as a superposition of two uncoupled  basis states $\ket{N,M_N,M_I}$
   \begin{equation}\label{eq:mixing}
  \ket{\gamma} =  c_{\gamma;1}  |N=1,M_N,M_I\rangle + c_{\gamma;2} |N=1,M_N',M_I'\rangle.
  \end{equation}

To obtain the mixing coefficients   $c_{7;1}$ and  $c_{7;2}$, consider the representation of molecular Hamiltonian \eqref{Hmol} in the  basis $\{ \ket{1,M_N,M_I}, \ket{1,M_N',M_I'}\}$ 
\begin{equation}\label{H2x2}
\begin{aligned}
H=\begin{pmatrix}
2 B_e +B_0\mu_I(g_IM_I+g_N M_N)+A M_N M_I  & A \sqrt{2-M_N(M_N+1)}
 \sqrt{3/4-M_I(M_I-1)}\\
A\sqrt{2-M’_N(M’_N-1)}  \sqrt{3/4-M’_I(M’_I+1)}& 2 B_e+B_0\mu_I(g_IM’_I+g_N M’_N)+A M’_N M’_I
\end{pmatrix}\\
\end{aligned}
\end{equation}
 The diagonal matrix elements are composed of the rotational, Zeeman and spin-rotation contributions, whereas the  off-diagonal elements are due to the  spin-rotation interaction.

 To be specific, we  choose to examine the magnetic field dependence of eigenstates $\ket{5}$ and $\ket{7}$ composed of the basis states $\ket{1,0,1/2}$ and $\ket{1,1,-1/2}$.
 In the basis of these states the matrix \eqref{H2x2} takes the form (in units of cm$^{-1}$)
\begin{equation}
\begin{aligned}
H=\begin{pmatrix}
3.675940  +B (0.0001785 )  & 8.209\times 10^{-7}  \\
8.209\times 10^{-7}    & 3.67594056  +B (-0.0001126) 
\end{pmatrix}\\
\end{aligned}
\end{equation}
At low magnetic fields, the diagonal matrix elements are nearly degenerate, and hence the uncoupled basis states are strongly mixed by the off-diagonal spin-rotation interaction terms.  As the field increases, the degeneracy is lifted and the eigenstates of the Hamiltonian become progressively closer to the bare uncoupled states $\ket{1,M_N,M_I}$.

We can see this more directly by looking at the eigenvectors of our model $2\times2$ Hamiltonian \eqref{H2x2} 
\begin{equation}
\begin{aligned}
|7\rangle=\cos({\theta}/{2}) |1,0,1/2\rangle+\sin({\theta}/{2}) |1,1,-1/2\rangle,\\
|5\rangle=-\sin({\theta}/{2}) |1,0,1/2\rangle+\cos({\theta}/{2}) |1,1,-1/2\rangle,\\
\tan{\theta}=\Big(\frac{\sqrt{2}\gamma}{B\mu (g_N-g_I)+\frac{\gamma}{2}}\Big),
\end{aligned}
\label{eq:mageigenvector}
\end{equation}
where the amplitudes $\cos(\frac{\theta}{2})$ and $\sin(\frac{\theta}{2})$ approximate the coefficients $c_{\gamma;1}$ and $c_{\gamma;2}$ in Eq.~\eqref{eq:mixing}. In the strong magnetic field limit, the argument $x=\sqrt{2}\gamma/(B\mu (g_N-g_I)+\frac{\gamma}{2})$ is small. Replacing  $\tan(x)\to x$, we obtain  $\theta\ \simeq C/B$, where $C=\sqrt{2}\gamma/\mu (g_N-g_I)$.

\begin{figure}[t!]
\begin{center}
\includegraphics[width=10cm]{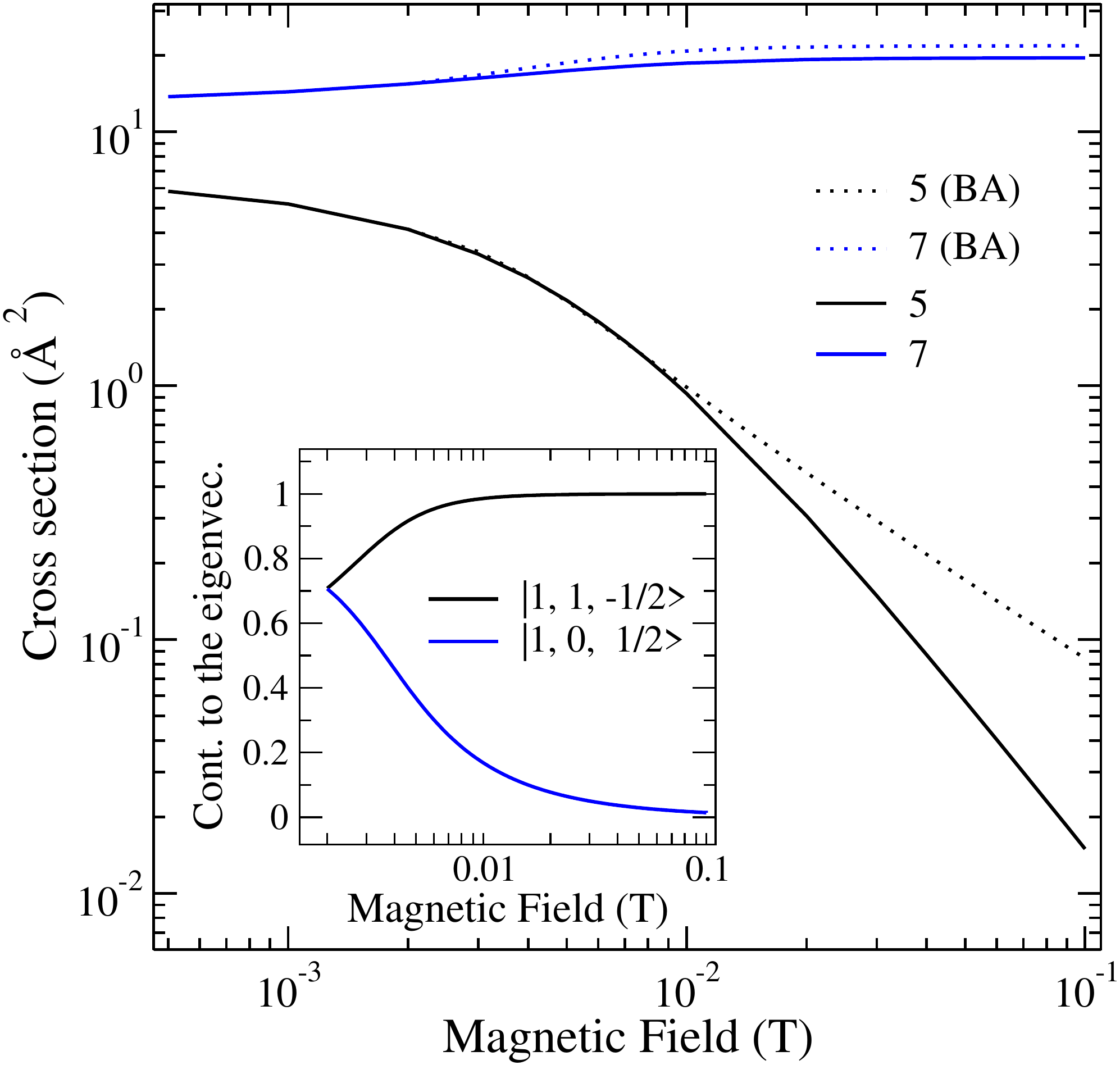}
\vspace{-0.7cm}
\end{center}
\caption{Magnetic field dependence of  He~+~CO  state-to-state cross sections  for the initial state  $\ket{8}$ and final states $\ket{5}$ and $\ket{7}$ at a collision energy of 2 K. Solid lines --   cross sections calculated using the accurate CC method. Dashed lines -- results obtained in the Born approximation (BA).  The insert shows the contribution of the basis states $\ket{N,M_N,M_I}$ to state $\ket{5}$  as a function of magnetic field.}
\label{fig:Crossfit}
\end{figure}

In the insert of Fig.~\ref{fig:Crossfit}, we plot the eigenvector components of the eigenstate $\ket{5}$  vs. magnetic field. We see that as the field increases, the spin-up component of state $\ket{5}$ decreases, and the eigenstate approaches the bare state  $\ket{1,1,-1/2}$. This numerical result  agrees with Eq.~(\ref{eq:mageigenvector}), which predicts $|5\rangle \to   |1,1,-1/2\rangle$  at large $B$, where  $\theta   \simeq C/B \to 0$.

Having discussed the magnetic field dependence of the Zeeman states, we can now  calculate the   inelastic cross sections between these states in the first Born approximation (here and throughout this work, we use atomic units, where $\hbar=1$)
\begin{equation}\label{eq:borncross_gen}
\sigma_{i\to j}^{B}= \frac{\mu^2}{4 \pi^2 } \left| \int e^{i \mathbf{q} \cdot \mathbf{R}} V_{ij}(\mathbf{R}) d^3R \right|^2,
\end{equation}
 where $\mathbf{q}$ is the transferred momentum, and $V_{ij}(\mathbf{R})$ is the matrix element of the interaction potential between the eigenstates $\ket{i}$ and $\ket{j}$.
Expanding our initial and final states in the bare state basis $\ket{N,M_N,M_I}$ defined above
\begin{equation}\label{eq:mageigenvector2}
\begin{aligned}
|i\rangle= c_{i1} |1,M_{N_i},1/2\rangle+c_{i2} |1,{M_{N_i}'},-1/2\rangle\\
|j\rangle= c_{j1} |1,M_{N_j},1/2\rangle+c_{j2} |1,{M_{N_j}’},-1/2\rangle.
\end{aligned}
\end{equation}
 the cross sections in the Born approximation (\ref{eq:borncross_gen}) can be written as
\begin{equation}\label{eq:borncross_ij}
\begin{aligned}
\sigma_{i\to j}^B= \frac{\mu^2}{4\pi^2} \bigg| \int e^{i \mathbf{q} \cdot \mathbf{R}} ( c_{i1} c_{j1} \langle 1,M_{N_j},1/2 |V|1,M_{N_i},1/2\rangle+c_{i1} c_{j2} \langle  1,{M_{N_j}’},-1/2  |V|1,M_{N_i},1/2\rangle \\
+c_{j1} c_{i2} \langle 1,M_{N_j},1/2 |V|1,{M_{N_i}'},-1/2\rangle+c_{i2} c_{j2} \langle  1,{M_{N_j}’},-1/2  |V|1,{M_{N_i}'},-1/2\rangle) d^3R \bigg|^2
\end{aligned}
\end{equation} 
The  atom-molecule interaction PES is independent of the nuclear spin, and hence diagonal in $M_I$. We can thus rewrite  Eq.~\eqref{eq:borncross_ij}  as
\begin{equation}\label{eq:borncross_ij2}
\begin{aligned}
\sigma_{i\to j}= \frac{\mu^2}{4 \pi^2} \bigg| \int e^{i \mathbf{q} \cdot \mathbf{R}} ( c_{i1} c_{j1} \langle 1,M_{N_j},1/2 |V|1,M_{N_i},1/2\rangle\\
+ c_{i2} c_{j2} \langle  1,{M_{N_j}’},-1/2  |V|1,{M_{N_i}'},-1/2\rangle) d^3R \bigg|^2.
\end{aligned}
\end{equation} 
For simplicity, we will consider the case of the fully spin-polarized initial Zeeman state such as  $\ket{8}$, where $c_{i2}=0$, and
  Eq.~\eqref{eq:borncross_ij2} simplifies to
\begin{equation}
\begin{aligned}
\sigma_{8\to 7}= \frac{\mu^2}{4 \pi^2 } \cos^2({\theta}/{2})  \left| \int e^{i \mathbf{q} \cdot \mathbf{R}} \langle 1, 1,1/2  |V|1, 0, 1/2 \rangle d^3R \right|^2,\\
\sigma_{8\to 5}= \frac{\mu^2}{4 \pi^2} \sin^2({\theta}/{2}) \left| \int e^{i \mathbf{q} \cdot \mathbf{R}} \langle1, 1,1/2  |V|1, 0,  1/2 \rangle d^3R \right|^2, \\
\end{aligned}
\label{eq:borncross}
\end{equation}
where $\sigma_{8\to 7}$ is the cross section for the nuclear spin-conserving transition $\ket{8} \to \ket{7}$ and $\sigma_{8\to 5}$ is the cross section for the nuclear spin-flipping transition  $\ket{8} \to \ket{5}$.
The prefactors $\cos^2({\theta}/{2})$ and $\sin^2({\theta}/{2})$ give the magnetic field scaling of the inelastic cross sections.

The cross sections calculated using Eq.~\eqref{eq:borncross} are compared with accurate CC calculations in Fig. \ref{fig:Crossfit}. 
The cross section $\sigma_{8\to 5}$ decreases with increasing magnetic field. This trend  is qualitatively reproduced by the Born approximation, which makes clear that the decline of the cross section $\sigma_{8\to 5}$ is caused by the decreasing mixing between the different nuclear spin components of the eigenstate $\ket{5}$ in Eq.~(\ref{eq:borncross}). As stated above, the mixing angle scales as $\theta\simeq B^{-1}$ in the large $B$-field limit, and thus  $\sigma_{8\to 5}\simeq B^{-2}$. 

In contrast, the cross section for the $\ket{8}\to \ket{7}$ transition tends to a constant  value in the high $B$-field limit due to the mixing coefficient  $\cos^2(\theta/2)$ in Eq.~(\ref{eq:borncross}) approaching unity as $\theta\to 0$. 
 At higher fields, the cross sections computed in the Born approximation start to deviate from the accurate CC results.
This could be a consequence of  multichannel effects, which become more pronounced  as the cross sections become smaller at higher $B$-fields.
The Born approximation does, however, capture  the overall  trend in the magnetic field dependence of scattering cross sections observed in  accurate CC calculations.

\subsection{Nuclear spin relaxation dynamics of CO molecules immersed in a cold He buffer gas}

In this section we examine the relaxation dynamics of CO molecules prepared in a single $N=0$ nuclear spin sublevel in cold collisions with He atoms. To this end, we use the standard rate  equations \cite{Ramakrishna:05,ramakrishna2006dissipative,blum2012density},
which describe the time evolution of populations of the individual molecular eigenstates $\ket{m}$
\begin{equation}\label{rate_eqs}
\begin{aligned}
\dot{\rho}_{mm}(t)=\sum_{n\neq m} \rho_{nn}(t) W_{mn}-\rho_{mm}(t) \sum_{n\neq m} W_{nm},
\end{aligned}
\end{equation}
where $W_{mn}=K_{n\to m}n_\text{He}$ are the  rates for the transitions $\ket{n}\to \ket{m}$  induced by collisions with helium atoms,  $n_\text{He}$ is the atomic density (we assume that $n_\text{He}\gg n_\text{CO}$), and $\rho_{nn}(t)$ is the diagonal density matrix element (the population of state $\ket{n}$) at the time $t$. The He~+~CO bimolecular transition rates  are obtained from  rigorous CC calculations as described above, and we take $n_\text{He}=10^{14}$~cm$^{-3}$.

\begin{figure}[t!]
\begin{center}
\includegraphics[width=12cm]{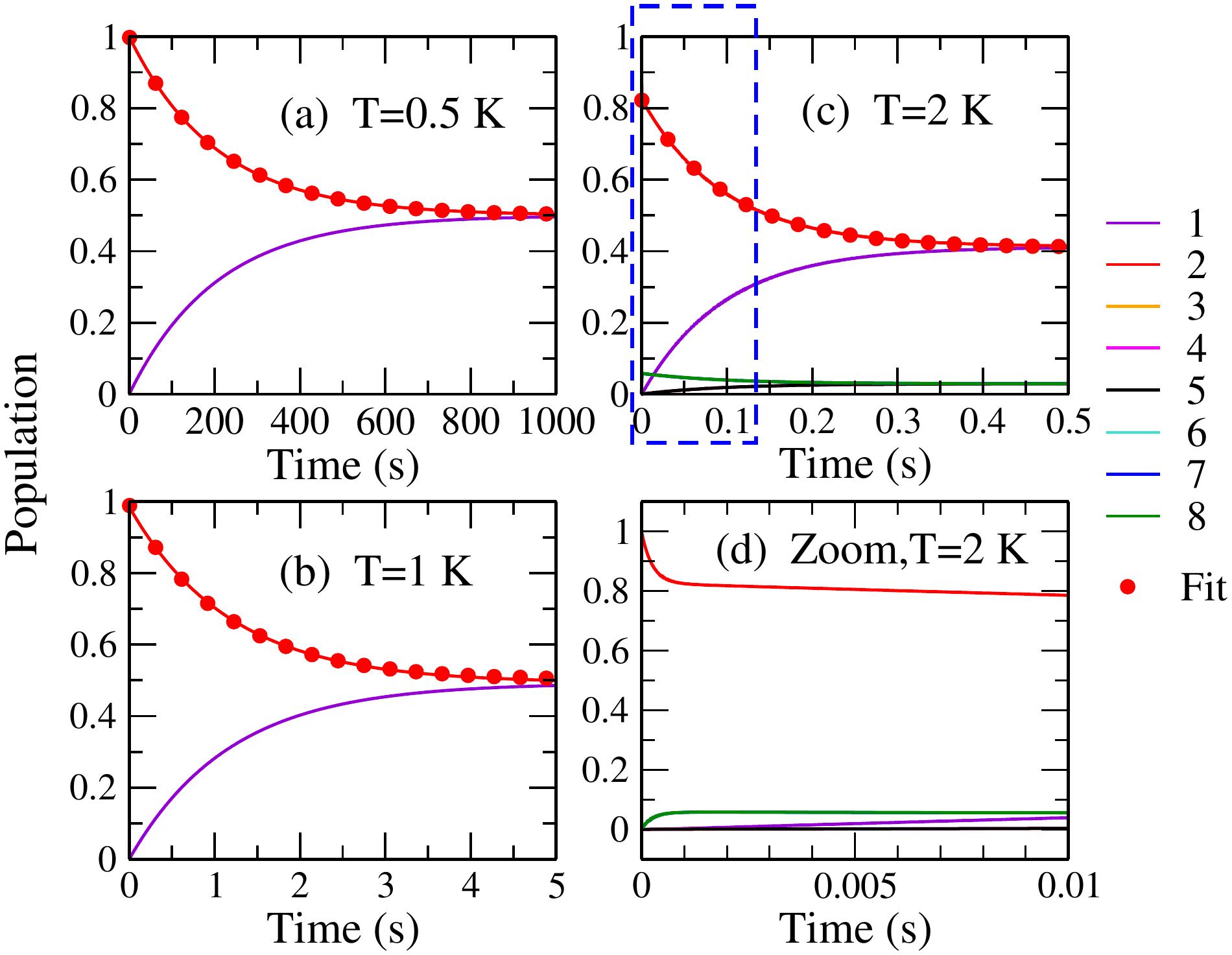}
\end{center}
\caption{ Time dynamics of CO nuclear spin state populations in cold collisions with He atoms at different temperatures: (a) $T=0.5$~K, (b) $T=1$~K and (c) $T=2$~K, respectively. Panel (d)  shows details of short-time dynamics at $T=2$~K. \textcolor{red}{The red dots denote} the exponential fit (see the text).}
\label{fig:time}
\end{figure}

Figure~\ref{fig:time} shows the time evolution of populations of the lowest nuclear spin states of CO molecules initially prepared in the pure nuclear spin state $\ket{2}$.  These results are obtained by solving the rate equations \eqref{rate_eqs} numerically using a basis set including the 8 lowest hyperfine states of CO (see Fig.~\ref{fig:elevels}).
The population of the initial state $\ket{2}$ relaxes to equilibrium due to collisions with buffer gas atoms. The equilibrium state is given by the Boltzmann distribution $\rho_{ij}^\text{eq}=\delta_{ij}e^{-E_i/k_B T}/Z$, where $Z$ is the partition function. At $T \ll 2B_e/k_B$, only the $N=0$ nuclear spin sublevels are thermally populated. The energy gap between the $N=0$ nuclear spin sublevels of CO is extremely small compared to $k_B T$ at $T\geq 1$~mK and $B < 0.1$~T [see Fig.~\ref{fig:elevels}], so we expect $\rho_{11}^\text{eq}=\rho_{22}^\text{eq}\simeq1/2$ as is indeed observed in Figs. \ref{fig:time}(a) and (b). At $T=2$~K a small fraction of the overall population ($<10\%$) ends up in $N=1$ rotational states as this temperature  is no longer negligibly small compared to the rotational spacing between the $N=0$ and $N=1$ levels (5.3~K).

The relaxation timescale is strongly temperature dependent. At $T=0.5$~K the relaxation is extremely slow, and thermal equilibrium is only reached on the timescale of hundreds of seconds. 
At higher temperatures the relaxation occurs much faster, taking $\simeq 0.5$~s at 2~K.
 Table~1 lists the nuclear spin relaxation times of CO molecules obtained by fitting the time dynamics of state populations in Fig.~\ref{fig:time} to an exponential form $\rho_{22}(t)=\frac{1}{2}(1+e^{-t/T_1})$. 

The drastic increase of the relaxation time of  state $\ket{2}$  at lower  temperatures can be explained by the extremely small transition rates for the nuclear spin changing transitions $\ket{2}\leftrightarrow \ket{1}$  (see Fig.~\ref{fig:rate12}), which lead to equilibration of the nuclear spin degrees of freedom.  The opposite trend is observed at higher temperatures,  where the relaxation times decrease dramatically due to the corresponding increase of the $\ket{2}\leftrightarrow \ket{1}$ transition rates.

\textcolor{red}{
We finally note that the nuclear spin relaxation times of CO molecules are moderately sensitive to the interaction PES (to within a factor of two), as shown in Appendix A.}

\begin{table}[ht]
\renewcommand{\arraystretch}{1.2} \addtolength{\tabcolsep}{5 pt}
\begin{center}
\begin{tabular}{cccc}
\hline \hline
$T(K)$ & 0.5 & 1.0 & 2.0 \\
$T_1$(s) &205&1.12&0.098\\     
\hline \hline
\end{tabular}
\caption{Nuclear spin relaxation times of CO molecules in the nuclear spin state $\ket{2}$ ($|N=0,M_I=1/2\rangle$) in collisions with He atoms. The magnetic field is 0.05~T.}
\end{center}
\label{tab:time}
\end{table}

\section{Summary and conclusions}


We have developed a rigorous CC methodology for quantum nuclear spin dynamics in cold, weakly anisotropic collisions between $^1\Sigma$ molecules and structureless atoms in the presence of an external magnetic field. As in all CC methods, the  solution of the time-independent Schr\"odinger equation is expanded  in a channel basis set. Here, we use a basis set composed of direct products of rotational and nuclear spin basis functions of the diatomic molecule. The theory is conceptually similar to the one developed previously for electron spin depolarization in cold atom-molecule collisions \cite{Volpi:02,Krems:04}, in which the electron spin basis functions are here replaced by their nuclear spin counterparts.
Nonetheless, our calculations show that nuclear  spin  relaxation occurs much more slowly than electron spin relaxation due to the much weaker interactions of nuclear spins with the rotational degrees of freedom.

We apply our methodology to study transitions between the different nuclear spin sublevels of $^{13}$C$^{16}$O in cold collisions with $^4$He buffer gas atoms. This system is experimentally relevant as buffer gas cooled  diatomic and polyatomic molecules have been  probed spectroscopically in a number of recent experiments \cite{Egorov:01,Lu:09,Patterson:13,Iwata:17,Changala:19,Santamaria:21,Hofsass:21,Daniel:21,Liu:22,Bu:22}.  We perform  rigorous coupled-channel quantum scattering calculations based on an accurate \textit{ab initio} potential energy surface of He-CO,  focusing on transitions between the nuclear spin sublevels of the ground ($N=0$) and the first excited ($N=1$) rotational states.


Our calculations show that state-to-state transitions between the nuclear spin sublevels of polar molecules (such as CO) in cold collisions with buffer-gas atoms (such as He)  are governed by several selection rules. The transition probability  depends on  (i) whether the  nuclear spin projection of the initial state changes in a collision and (ii) whether the initial and final states are fully spin-polarized.
 The dominant  transitions  that belong to group I conserve the nuclear spin projection $M_I$ but  change either $N$ or $M_N$. Group II transitions, which change $M_I$ and occur between non fully polarized initial and/or final Zeeman states, are three orders of magnitude slower. Finally, the weakest group-III transitions change $M_I$ and occur between the fully polarized initial and final Zeeman states. 
 \textcolor{red}{This hierarchy of transitions is expected:  based on the weakness of the nuclear spin-rotation coupling between the different rotational-Zeeman levels, one could predict that the strongest transitions would be those that conserve $M_I$. In addition, similar propensity rules for electron spin-rotational transitions were found in our previous work on cold $^4$He + $^{40}$CaH collisions\cite{Koyu:22}.
However, because the rotational-Zeeman levels are coupled in a non-trivial way by the anisotropy of the atom-molecule interaction potential, only rigorous quantum scattering calculations can provide quantitative insight into the hierarchy of  transitions.} 

We find that only group II transitions have a marked magnetic field dependence. The origin of this dependence is the field-induced mixing between the different spin-rotational basis functions that compose the initial and final molecular eigenstates.
Specifically, as the magnetic field increases, the eigenstates become more polarized, i.e., they acquire a definite value of the nuclear spin projection $M_I$. Because the atom-molecule interaction potential is diagonal in $M_I$, both the initial and final states must have the same value of $M_I$ for the transition to occur in the first order. As a result, the cross sections for group-II transitions originating from the fully spin-polarized eigenstates, are proportional to the overlap between the nuclear spin components of the initial and final states. This overlap is most strongly field-dependent when the initial and final states have the magnetic $g$-factors of the opposite sign, such as the states $\ket{8}$ and $\ket{5}$ in Fig.~\ref{fig:elevels}. In this case, the overlap scales as $1/B$, and the inelastic cross section as $1/B^2$.

In contrast, for group-I transitions the nuclear spin overlap factor depends on the magnetic field only weakly since both the initial and final states have the same $M_I$ values. For group-III transitions, the nuclear spin overlap is zero at all magnetic fields, and transitions occur through an indirect mechanism involving field-independent couplings to rotationally excited states. We observe a number of Feshbach resonances in the collision energy dependence of the integral cross sections, which are particularly pronounced  for group-III transitions.

Finally, we explored the time dynamics of nuclear spin relaxation of CO molecules immersed in a cold gas of He atoms using rate equation simulations based on the CC collision rates computed in this work.
While the relaxation times of the $N=0$ nuclear spin sublevels   are extremely long at very low temperatures ($T\ll 2B_e/k_B$), they 
decline sharply with increasing temperature due to a dramatic increase in the  nuclear spin flipping rates.
Our simulations thus indicate that preparing long-lived nuclear spin sublevels of diatomic molecules in inert buffer gases would require cooling the molecules to temperatures much lower than the spacing between the $N=0$ and $N=1$ rotational states.

 \textcolor{red}{Because spin-flipping transitions in the $N=0$ manifold belong to  group III (and hence their cross sections are extremely small), the extremely long relaxation times  might be expected. However,  the precise values of collision rates responsible for the nuclear spin relaxation timescales  cannot be determined  without performing the rigorous quantum scattering calculations reported here.} 



\section*{Acknowledgements}
We thank Jonathan Weinstein for stimulating discussions, and Balakrishnan Naduvalath for providing the results of quantum dynamics calculations of cold $^4$He~+~$^{12}$C$^{16}$O  collisions \cite{Balakrishnan2000jcp,Yang2005jcp}. This work was supported by the NSF through the CAREER program (PHY-2045681).

\section*{\textcolor{red}{Appendix A: Sensitivity to  the interaction potential}}
\textcolor{red}{In this section we explore the sensitivity of $^4$He~+~$^{13}$C$^{16}$O collision  cross sections, rate constants, and relaxation times to small changes in the interaction potential. This is necessary because low-temperature scattering observables are known to be strongly affected by such changes \cite{Balakrishnan:16,Morita:19b}, and the  accuracy of {\it ab initio} interaction potentials, such as the He-CO potential used in this work, is limited.
To vary the interaction potential, we multiply it by a constant scaling factor $\lambda$ as done in previous theoretical work \cite{Morita:19b}. As the He-CO scattering calculations are computationally intensive,  we choose two values of the scaling parameter, $\lambda=1.02$ and $\lambda=0.98$, to explore the range of   uncertainty in the He-CO interaction potential of $\pm 2\%$.}

\begin{figure}[t]
 \centering
     \includegraphics[width=14cm]{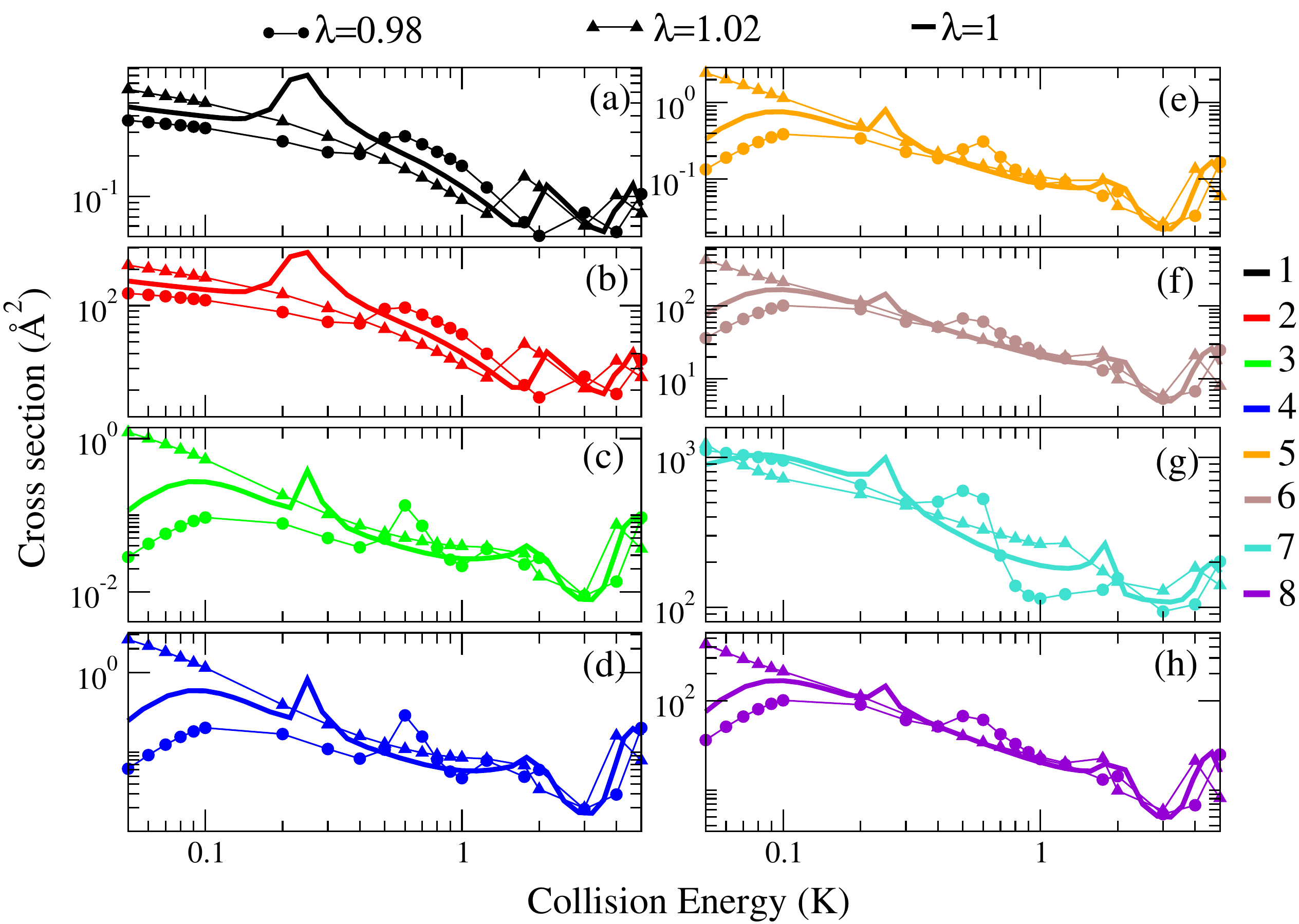}
 \caption{\textcolor{red}{State-to-state cross sections for He~+~CO collisions from the initial state $\ket{7}$ as a function of collision energy. Each panel represents a different final state starting from $\ket{1}$ and ending with $\ket{8}$. The magnetic field is 0.05 T.}
}
  \label{fig:CrossSectionpot7}
\end{figure}

\begin{figure}[t]
 \centering
     \includegraphics[width=14cm]{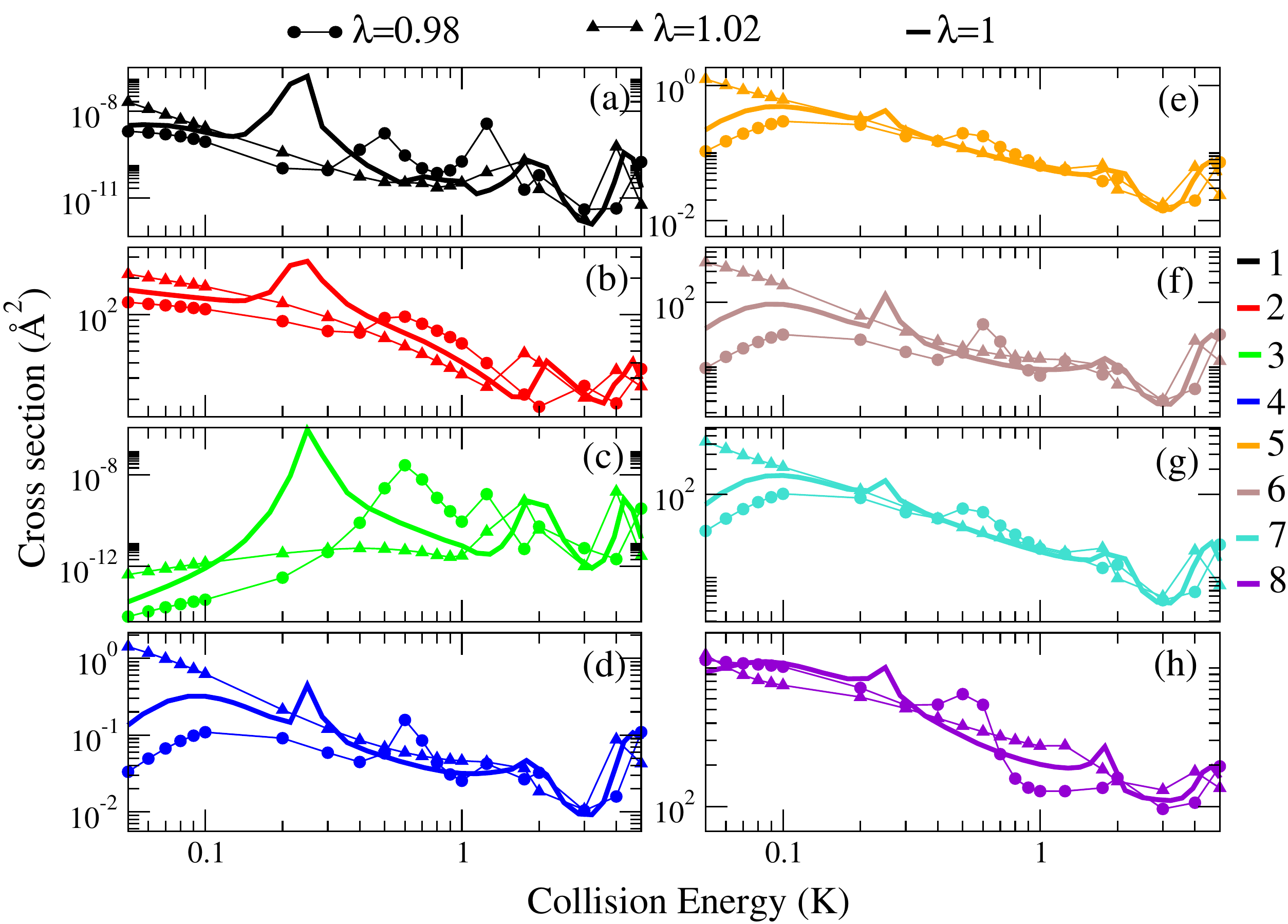}
 \caption{\textcolor{red}{State-to-state cross sections for He~+~CO collisions from the initial state $\ket{8}$ as a function of collision energy. Each panel represents a different final state starting from $\ket{1}$ and ending with $\ket{8}$. The magnetic field is 0.05 T.
}}
  \label{fig:CrossSectionpot8}
\end{figure}

\color{red}
Figures~\ref{fig:CrossSectionpot7} and \ref{fig:CrossSectionpot8} show the cross sections for the initial states $\ket{7}$ and $\ket{8}$ to all final states calculated for the unscaled ($\lambda=1$) and scaled interaction potentials.
We observe that altering the potential changes the background values of scattering cross sections and shifts the positions of scattering resonances. This is  expected as the scaling changes the positions of the last bound states of the He-CO collision complex, which are responsible for Feshbach resonances.   The changes are particularly pronounced in the s-wave regime below 0.1~K.



Table 2 lists the state-to-state He-CO collision rates calculated with unscaled $(\lambda=1)$ and scaled $(\lambda=0.98,\, 1.02$) interaction PESs.
While the largest rates do not change significantly upon PES scaling, the smaller rates that correspond to group-II and group-III transitions are more sensitive to the PES. The most sensitive are group-III transitions, whose rates  are the smallest, and most affected by scattering resonances.

\begin{table}[ht]
\renewcommand{\arraystretch}{1.2} \addtolength{\tabcolsep}{5 pt}
\begin{center}
\begin{tabular}{cccc}
\hline \hline
transition & +2\% & -2\% & unchanged\\
\hline    
$|1\rangle \to |2\rangle$ &$5.45\times 10^{-23}$&$3.26\times 10^{-24}$&$6.69\times 10^{-23}$\\    
$|1\rangle \to |8\rangle$ &$8.06\times 10^{-27}$&$1.23\times 10^{-26}$&$5.10\times 10^{-26}$\\    
$|2\rangle \to |1\rangle$ &$5.45\times 10^{-23}$&$3.27\times 10^{-24}$&$6.37\times 10^{-23}$\\    
$|2\rangle \to |3\rangle$ &$8.06\times 10^{-27}$&$1.24\times 10^{-26}$&$5.10\times 10^{-26}$\\   
$|3\rangle \to |2\rangle$ &$2.38\times 10^{-22}$&$3.65\times 10^{-22}$&$2.00\times 10^{-21}$\\    
$|3 \rangle \to |8\rangle$ &$2.24\times 10^{-23}$&$1.66\times 10^{-21}$&$1.25\times 10^{-20}$\\
$|6\rangle  \to |3\rangle$ &$1.05\times 10^{-13}$&$9.01\times 10^{-14}$&$7.01\times 10^{-14}$\\   
$|8 \rangle \to |1\rangle$ &$2.38\times 10^{-22}$&$3.64\times 10^{-22}$&$2.00\times 10^{-21}$\\     
$|8\rangle \to |2\rangle$ &$3.72\times 10^{-11}$&$3.95\times 10^{-11}$&$3.81\times 10^{-11}$\\    
$|8\rangle \to |3\rangle$ &$2.23\times 10^{-23}$&$1.66\times 10^{-21}$&$1.25\times 10^{-20}$\\    
$|8\rangle \to |6\rangle$ &$1.92\times 10^{-11}$&$9.97\times 10^{-12}$&$8.89\times 10^{-12}$\\    
\hline \hline
\end{tabular}
\caption{\textcolor{red}{State-to-state He~+~CO collision rates  calculated for the unscaled PES (the last column) and scaled PESs (the first and  second columns). These transitions where calculated at 0.5 K. The magnetic field is 0.05~T. }}
\end{center}
\label{tab:rates}
\end{table}

\begin{figure}
 \centering
     \includegraphics[width=12cm]{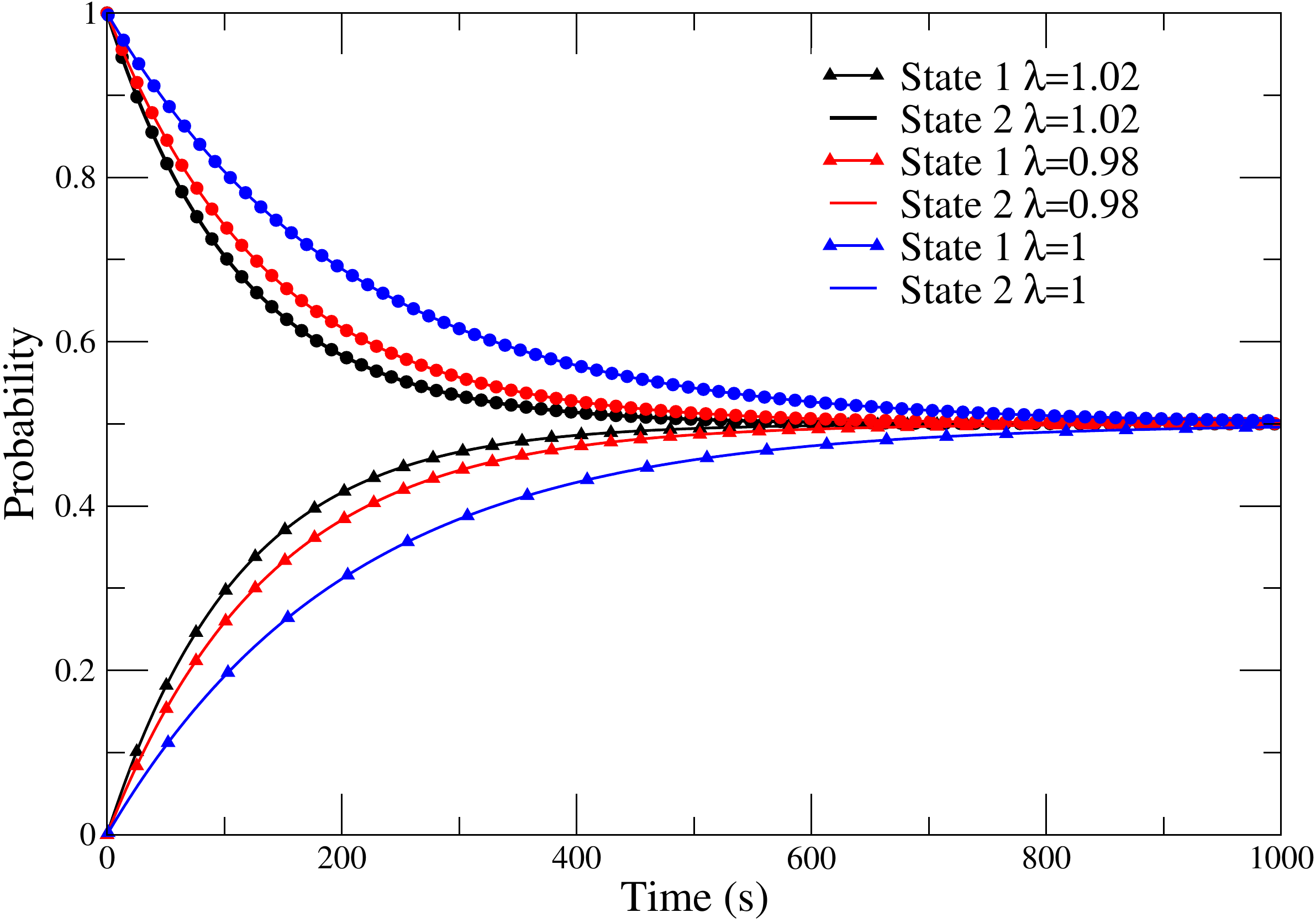}
 \caption{\textcolor{red}{Time dynamics of the nuclear spin state populations of CO in cold collisions with He atoms at $T = 0.5$~K for the unscaled and scaled He-CO interaction potentials.  The dots denote the exponential fits, which provide the $T_1$ times.}}
  \label{fig:T1changenPot}
\end{figure}

Lastly, we explore the sensitivity of nuclear spin relaxation dynamics to small changes in the He-CO interaction PES.  Figure~\ref{fig:T1changenPot} shows the time evolution of the populations of the lowest nuclear spin states of CO  initially prepared in the pure nuclear spin state $|2\rangle$ for unscaled $(\lambda=0)$ and scaled ($\lambda=0.98, \, 1.02$)  interaction potentials. 
We observe that although changing the interaction potential has an impact on the details of relaxation dynamics, the qualitative features of the dynamics are not affected.
 This is because the relaxation timescales are determined by many state-to-state transitions, and the overall effect of the interaction PES on these transition rates tends to average out.
 The nuclear spin relaxation times of CO molecules in a buffer gas of He listed in Table 3 are seen to change by a factor of two when the interaction PES is varied by $\pm 2\%$.
 
 

\color{black}

\begin{table}[ht]
\renewcommand{\arraystretch}{1.2} \addtolength{\tabcolsep}{5 pt}
\begin{center}
\begin{tabular}{cccc}
\hline \hline
PES type & Unscaled & $\lambda=1.02$ & $\lambda=0.98$ \\
\hline
$T_1$(s) &205&111&137\\     
\hline \hline
\end{tabular}
\caption{\textcolor{red}{Nuclear spin relaxation times of CO molecules in the nuclear spin state $\ket{2}$ ($|N=0,M_I=1/2\rangle$) in collisions with He atoms at $T=0.5$~ K. The magnetic field is 0.05~T. }}
\end{center}
\label{tab:time2}
\end{table}

\section*{Appendix \textcolor{red}{B}: Comparison with previous He~+~CO calculations}

\begin{figure}[t]
 \centering
 \includegraphics[width=11cm]{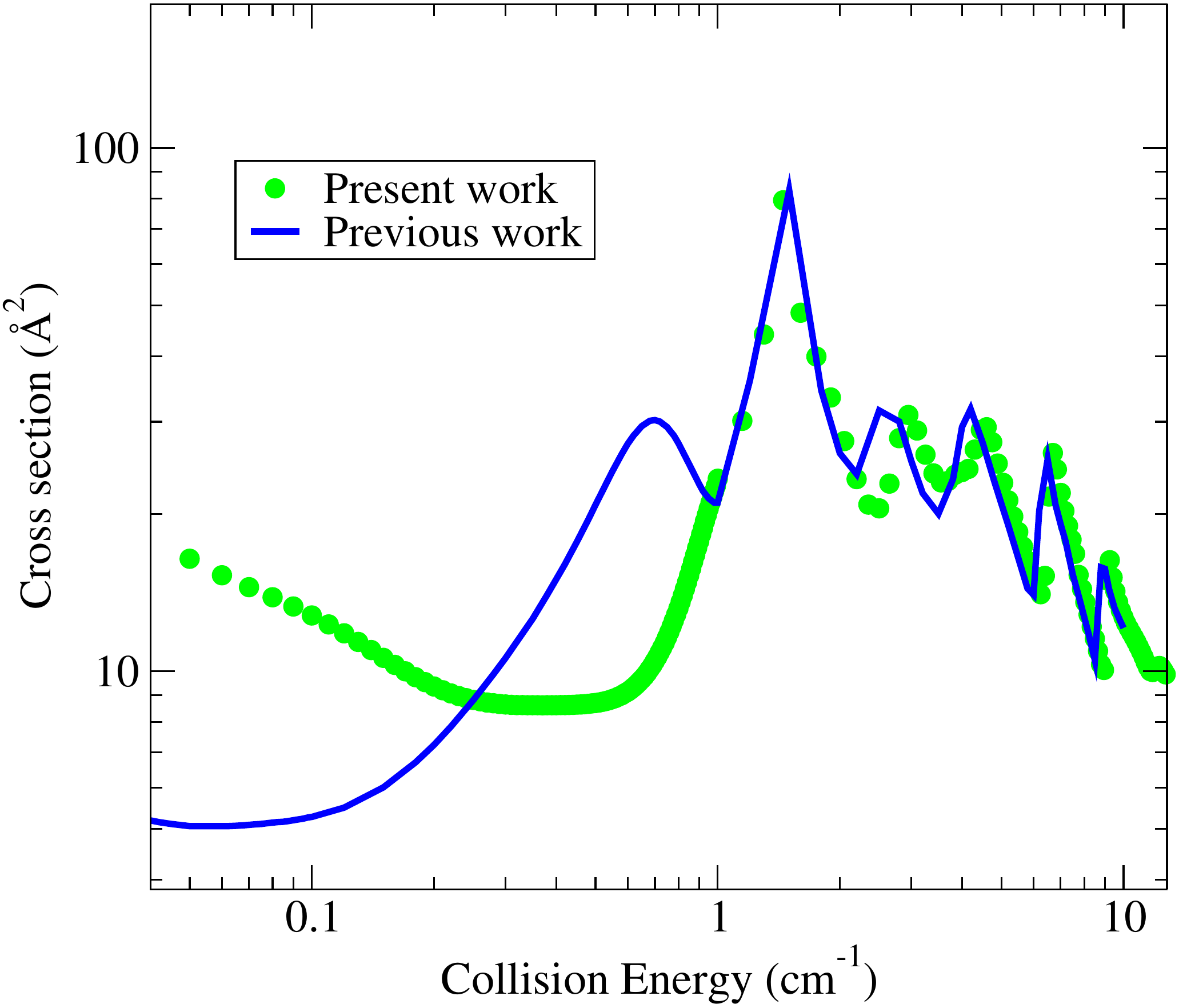}
 \caption{Collision energy  dependence of  the inelastic cross section for the $N=1\to 0$ transition in $^4$He~+~$^{12}$C$^{16}$O collisions calculated using the present code (green/grey dots). The results of independent quantum scattering calculations \cite{Balakrishnan2000jcp,Yang2005jcp} are shown as the solid blue (grey) line.}
  \label{fig:test}
\end{figure}

To test our newly developed quantum scattering code, we calculated the cross sections for rotational relaxation in $^4$He~+~$^{12}$C$^{16}$O collisions as a function of collision energy. The results are compared in Fig.~\ref{fig:test} with the reference calculations performed by Balakrishnan {\it et al.} \cite{Balakrishnan2000jcp} and by Yang {\it et al.} \cite{Yang2005jcp}. Note that the $^{12}$C$^{16}$O isotope lacks the nuclear spin, so these calculations (unlike those on $^4$He~+~$^{13}$C$^{16}$O) do not account for it.
 In order to compare with the previous results computed using a total  angular momentum basis \cite{Balakrishnan2000jcp,Yang2005jcp}, we  averaged our $M_N$-resolved cross sections over  the three degenerate $M_N$ components of the $N=1$ initial state.

As shown in Fig.~\ref{fig:test}, our cross sections are in good agreement with the reference values at collision energies above 1 cm$^{-1}$. In particular, the positions and widths of  five scattering resonances, which occur between 1 and 10 cm$^{-1}$, agree closely. We observe a significant discrepancy at the lowest-energy resonance at $E=0.7$ cm$^{-1}$ and in the $s$-wave threshold regime, where our results are above the reference values.
We attribute these discrepancies to  small differences in the rotational constants of CO, and in the reduced masses of He-CO used in the present and the previous \cite{Balakrishnan2000jcp,Yang2005jcp} calculations. It is well established that such small differences can have a large effect on scattering observables at ultralow temperatures \cite{Balakrishnan:16,Morita:19b}. 

\section*{Appendix \textcolor{red}{C}: PES scaling}

Here, we describe the scaling of the He-CO PES, which is necessary to account for the shift of the center of mass of $^{13}$CO compared to $^{12}$CO.
To this end we follow the procedure described in Ref.~\citenum{Zuchowski:09}.
The  transformation between the He-$^{13}$C$^{16}$O Jacobi coordinates $(R, r, \theta)$ and the He-$^{12}$C$^{16}$O Jacobi coordinates $R', r', \theta'$ is given by (assuming $r=r'$)
\begin{equation}
\begin{aligned}
R&=R'\sqrt{1+(\Delta/R')^2-2 \Delta/R' \cos\theta'}, \\
\theta&=\arccos\left(\dfrac{R'\cos\theta'-\Delta}{R}\right), 
\end{aligned}
\end{equation}
where $\Delta$ is the shift of the center of mass position between $^{13}$C$^{16}$O and $^{12}$C$^{16}$O.

\section*{Appendix \textcolor{red}{D}: Magnetic field dependence of  He~+~CO cross sections for the initial states $\ket{3}$, $\ket{4}$, $\ket{5}$ and $\ket{6}$.}

\begin{figure}[t]
 \centering
 \includegraphics[width=11cm]{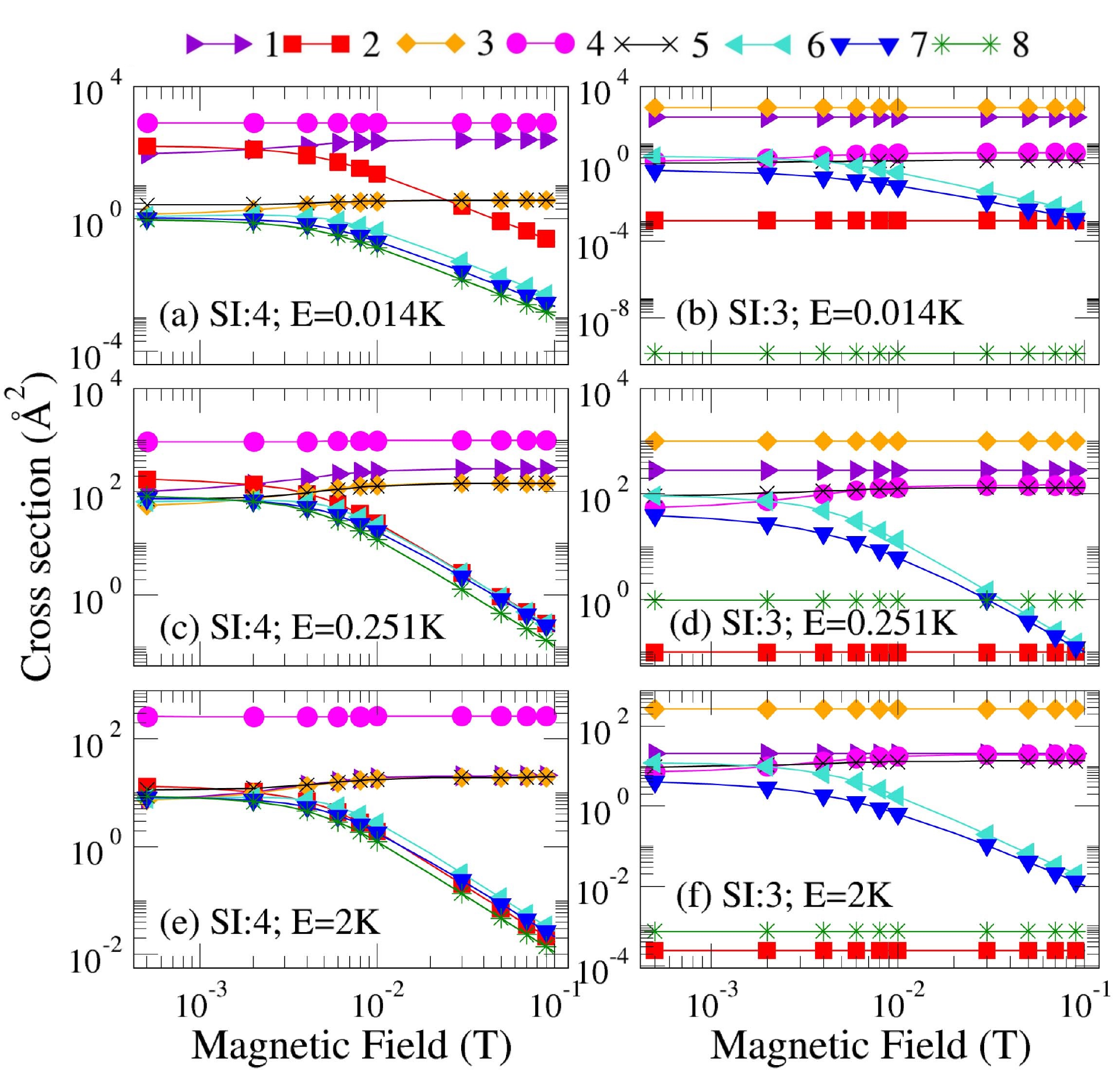}
 \caption{Magnetic field dependence of state-to-state cross sections for   $^4$He~+~$^{13}$C$^{16}$O collisions. The collision energy and the initial state index are indicated in each panel. The cross sections $\sigma_{3\to 2}$ and $\sigma_{3\to 8}$ are multiplied by $10^6$ to aid visibility. \textcolor{red}{The various color-coded symbols indicate the final state indices (SIs) specified in Fig.~\ref{fig:elevels}.}}
  \label{fig:M34_sup}
\end{figure}

\begin{figure}[t]
 \centering
 \includegraphics[width=11cm]{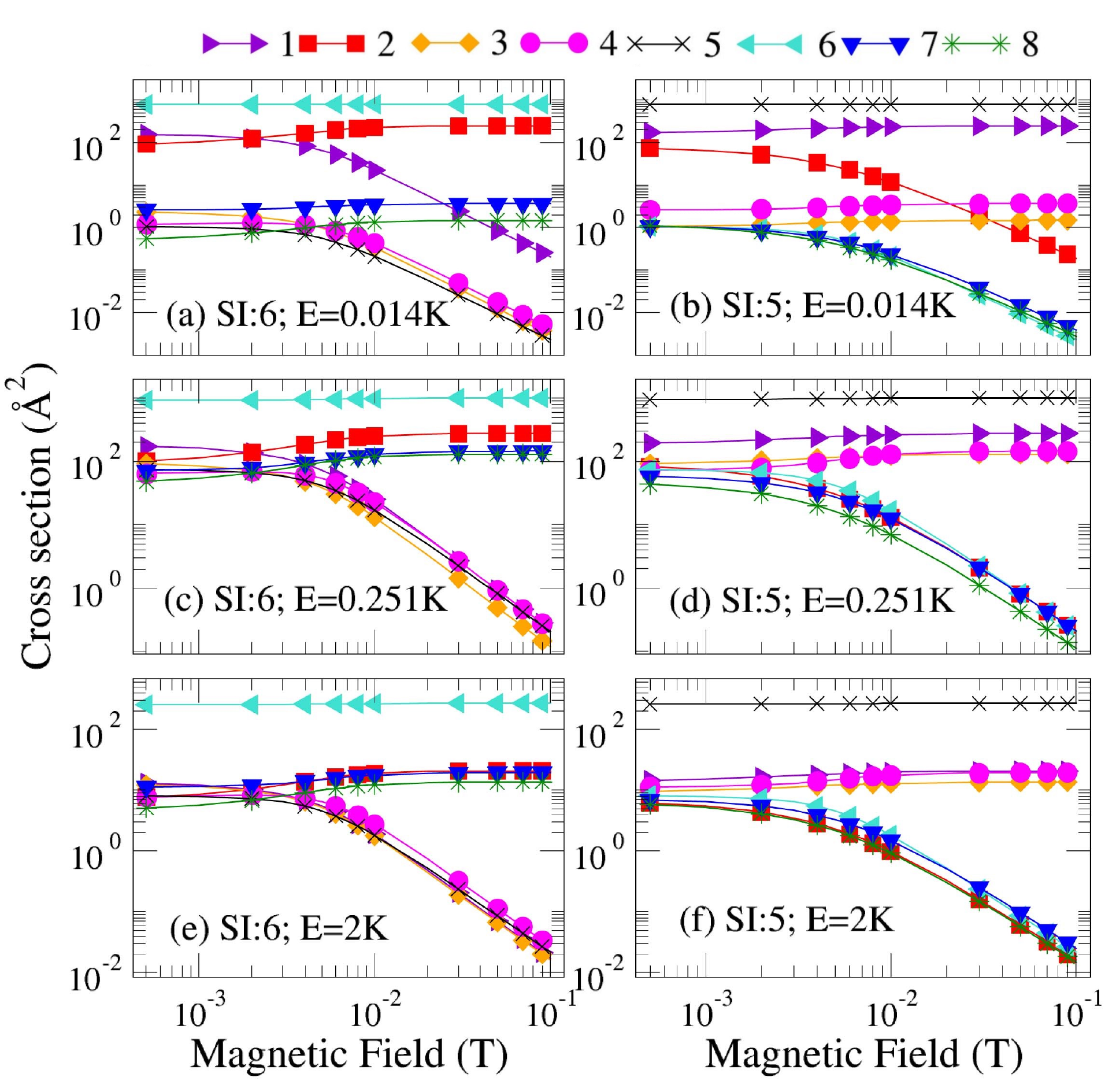}
 \caption{Magnetic field dependence of state-to-state cross sections for $^4$He~+~$^{13}$C$^{16}$O collisions. The collision energy and the initial state index  are indicated in each panel. \textcolor{red}{The various color-coded symbols indicate the final state indices (SIs) specified in Fig.~\ref{fig:elevels}.}}
  \label{fig:M56_sup}
\end{figure}

The magnetic field dependence of the cross sections for the initial states $\ket{1}$, $\ket{2}$, $\ket{7}$, and $\ket{8}$ is discussed in the main text (see section III C). Here, we present additional results for the initial states $\ket{3}$, $\ket{4}$, $\ket{5}$ and $\ket{6}$ (see Figs.~\ref{fig:M34_sup} and \ref{fig:M56_sup}).

In Figs.~\ref{fig:M34_sup}(b), (d), and (f),  we observe three groups of transitions for the fully spin-stretched initial state $\ket{3}$ in the $N=1$ manifold.
As discussed in the main text, nuclear spin-conserving (group I) transitions to the final states $\ket{1}$, $\ket{4}$ and $\ket{5}$ are magnetic field-independent. The nuclear spin-flipping transitions involving the fully spin-polarized final states $\ket{2}$ and $\ket{8}$ belong to group III and have the smallest cross sections, which show no magnetic field dependence  as well. Only group II transitions have a significant magnetic field dependence. 

The initial state $\ket{4}$ is not fully spin polarized, so transitions out of this state can belong to either  group I or group II.  In panels (a), (c), and (e) of Fig.~\ref{fig:M34_sup}, we show the cross  sections for these transitions.  As before, nuclear spin-conserving transitions to final states $\ket{1}$, $\ket{3}$ and $\ket{5}$ (group-I transitions) are independent of magnetic field. In contrast, nuclear spin-flipping transitions from group II populating the final states $\ket{2}$, $\ket{6}$, $\ket{7}$ and $\ket{8}$ display a strong magnetic field dependence.

Figure~\ref{fig:M56_sup} shows the cross sections for the initial states $\ket{5}$ and $\ket{6}$. We observe very similar  trends to those discussed above for the initial states $\ket{3}$ and $\ket{4}$.  The transitions can be classified in two groups:  Group I ($\ket{5} \to \ket{1}$, $\ket{3}$, $\ket{4}$; 
                   $\ket{6} \to \ket{2}$, $\ket{7}$, $\ket{8}$) and group II ($\ket{5} \to \ket{2}$, $\ket{6}$, $\ket{7}$, $\ket{8}$; $\ket{6} \to \ket{1}$, $\ket{3}$, $\ket{4}$, $\ket{5}$).

\bibliographystyle{unsrt}

\bibliography{COHe, cold_mol}
\end{document}


\title{Nuclear spin depolarization in cold atom-molecule collisions }

\author{Rebekah Hermsmeier}
\affiliation{Department of Physics, University of Nevada, Reno, NV 89557, USA}

\author{Xiaodong Xing}
\affiliation{Department of Physics, University of Nevada, Reno, NV 89557, USA}

\author{Timur V. Tscherbul}
\affiliation{Department of Physics, University of Nevada, Reno, NV 89557, USA}

\maketitle

\section{Magnetic field dependence of the state-to-state He~+~CO cross sections for initial states: $\ket{3}$, $\ket{4}$, $\ket{5}$ and $\ket{6
}$.}

\begin{figure}[]
 \centering
 \includegraphics[width=11cm]{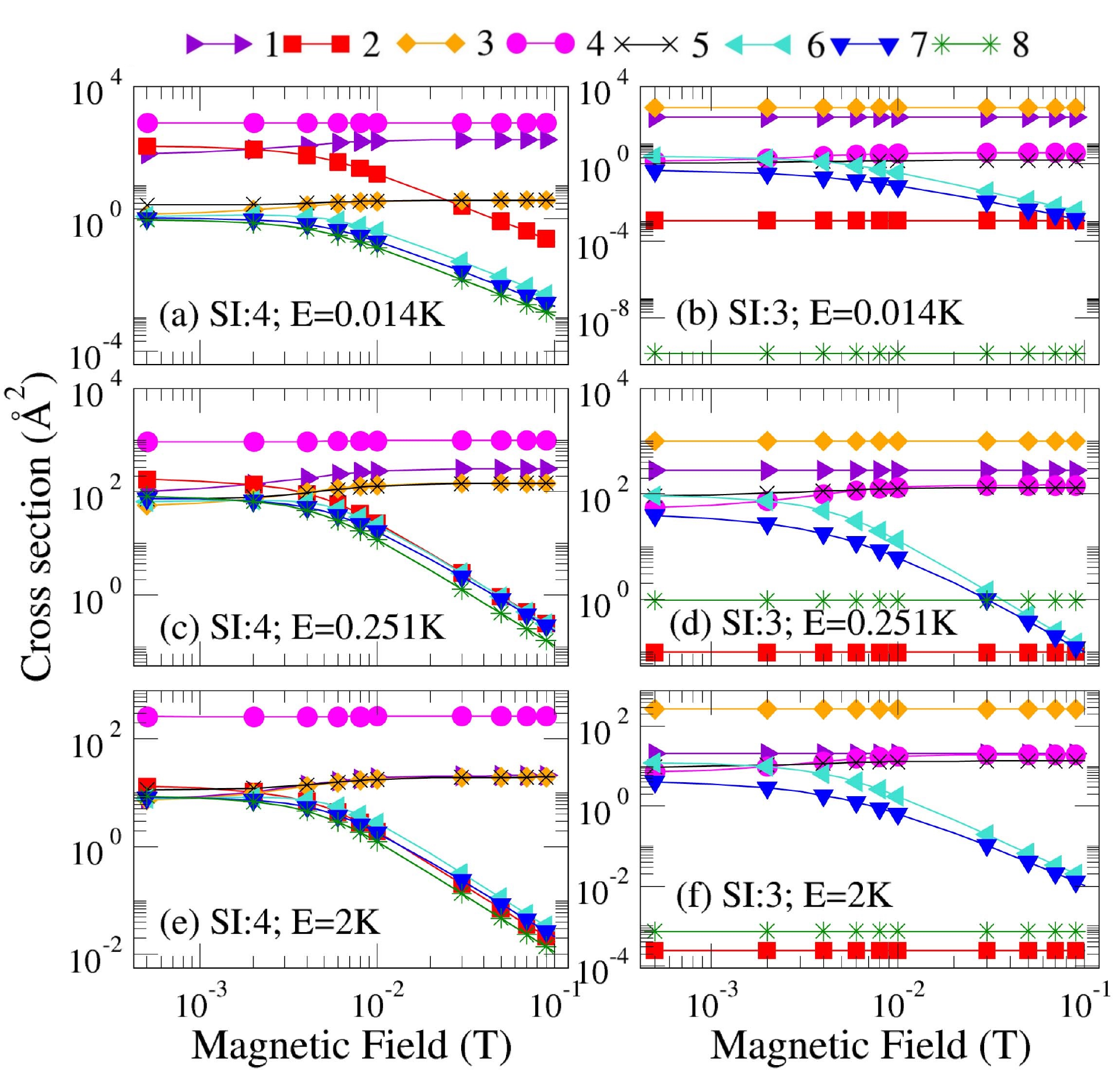}
 \caption{Magnetic field dependence of state-to-state cross sections for  $^{13}$C$^{16}$O$-^4$He collisions. The collision energy and the initial SI are indicated in each panel. The order of magnitude of the cross section for transitions SI:3 $\to$ SI:2 and SI:3 $\to$ SI:8 has been multiplied by $10^6$ for the sake of sizability.}
  \label{fig:M34_sup}
\end{figure}

\begin{figure}[]
 \centering
 \includegraphics[width=10cm]{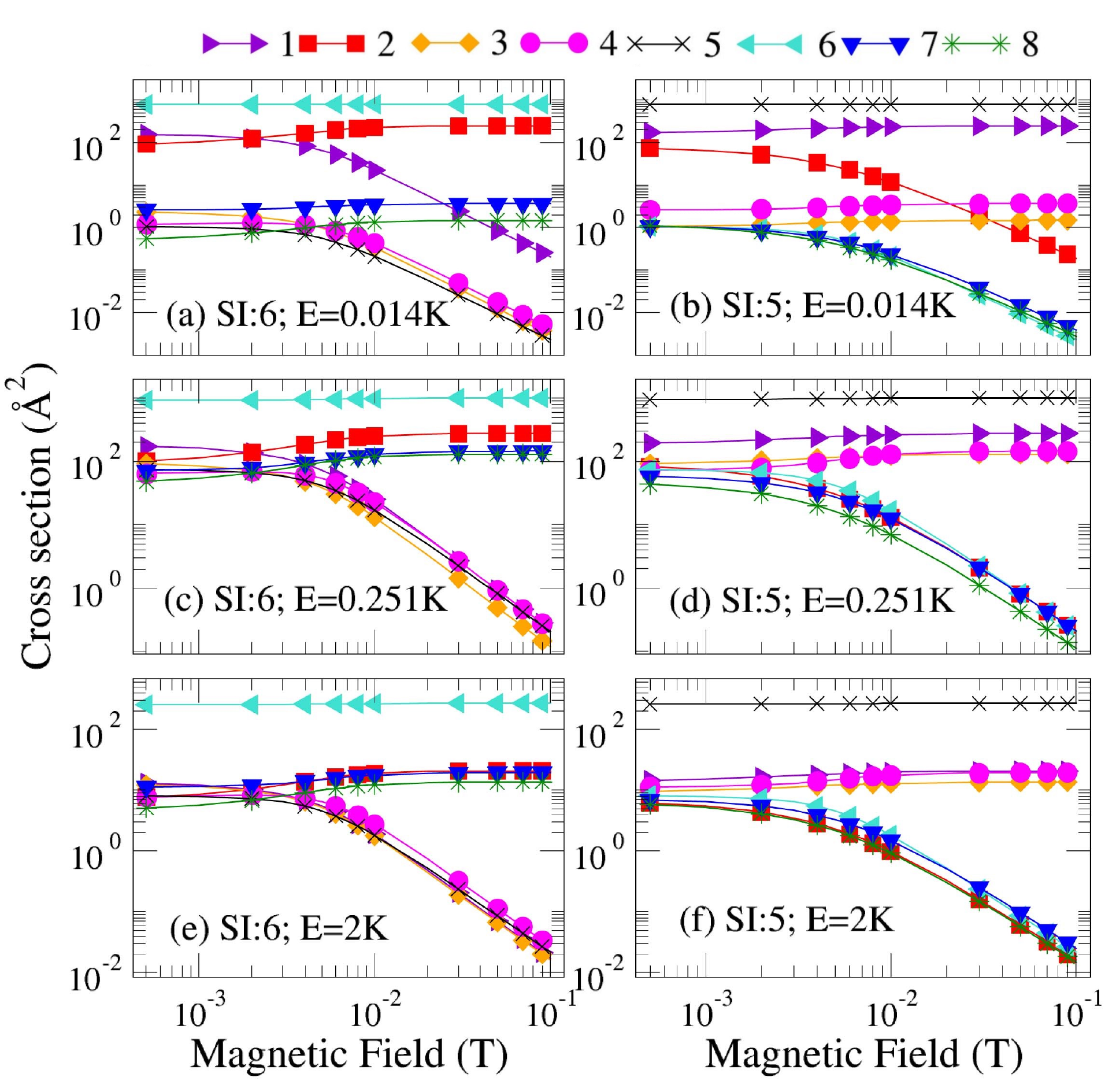}
 \caption{Magnetic field dependance of state-to-state cross sections for  $^{13}$C$^{16}$O$-^4$He collisions. The collision energy and the initial SI are indicated in each panel. }
  \label{fig:M56_sup}
\end{figure}

We discussed the features of magnetic field dependence of transition cross sections for initial states $\ket{1}$, $\ket{2}$, $\ket{7}$ and $\ket{8}$ in the main text (see section III-C). Here, we present the results in the same fashion with respect to the initial states $\ket{3}$, $\ket{4}$, $\ket{5}$ and $\ket{6
}$  in Fig.~\ref{fig:M34_sup} and Fig.~\ref{fig:M56_sup}.

In the rotational manifold ($N=1$), we observe the spin stretched initial state $\ket{3}$ ($s=1/2$) undergoes three transition pathways as displayed in panels (b), (d), and (f) of Fig.~\ref{fig:M34_sup}. Group I transitions have larger cross sections and admit the nuclear spin projection conserved final states $\ket{1}$, $\ket{4}$ and $\ket{5}$, which display an independence on magnetic field. The spin-flipping transitions (group III) involving spin polarized final states $\ket{2}$ and $\ket{8}$ ($s=-1/2$) have the smallest cross section and show no dependence on magnetic field as well. Only the group II transitions yield significant magnetic field dependence characteristics. Compare with initial state $\ket{3}$, the corresponding final states $\ket{6}$ and $\ket{7}$ have opposite nuclear spin  projections and same rotational quantum number $N=1$. 

In terms of initial state $\ket{4}$ (non spin stretched state), transitions can be cataloged by final states into group I and group II in panels (a), (c), and (e) of Fig.~\ref{fig:M34_sup}. The nuclear spin conserved transitions into final states $\ket{1}$, $\ket{3}$ and $\ket{5}$ show out independence of magnetic field and belonges to the transition group I. The strong magnetic field dependence are found in the transitions of final states $\ket{2}$ for  $N=0$ as well as $\ket{6}$, $\ket{7}$ and $\ket{8}$ for $N=1$. All of these final states also have different sign of nuclear spin comparing to the initial $\ket{4}$.

Similarly, we can discuss  the findings of initial states $\ket{5}$ and $\ket{6}$ with the same arguments. To avoid the repetition, we specify the transitions presented in Fig.~\ref{fig:M56_sup}, as follows:
\begin{itemize}
    \item Group I: $\ket{5} \to \ket{1}$, $\ket{3}$, $\ket{4}$; 
                   $\ket{6} \to \ket{2}$, $\ket{7}$, $\ket{8}$
    \item Group II: $\ket{5} \to \ket{2}$, $\ket{6}$, $\ket{7}$, $\ket{8}$; 
                   $\ket{6} \to \ket{1}$, $\ket{3}$, $\ket{4}$, $\ket{5}$ 
\end{itemize}
It should be noted that group III transitions do not happen for both initial $\ket{5}$ and $\ket{6}$ states.

